\documentclass[acmsmall]{acmart}

\acmJournal{PACMPL}
\acmYear{2020} \acmVolume{4} \acmNumber{ICFP} \acmArticle{105} \acmMonth{8} \acmPrice{}\acmDOI{10.1145/3408987}

\setcopyright{none}

\usepackage{booktabs}   
\usepackage{subcaption} 

\usepackage[T1]{fontenc}
\usepackage{flushend}
\usepackage[scaled=0.75]{beramono}
\usepackage{amsmath,stmaryrd}
\usepackage[inference]{semantic}
\usepackage{xspace}
\usepackage{multirow}
\usepackage{mathtools}
\usepackage{pbox}
\usepackage{paralist}
\usepackage{fixltx2e}
\usepackage{listings,atbegshi,ifthen}
\usepackage{ifplatform}
\iflinux
 \input{lstlinebgrdfixmepls}
\fi
\usepackage{lstlinebgrd}
\usepackage{algorithm}
\usepackage{hyperref}
\usepackage[noend]{algpseudocode}
\usepackage{resizegather}
\usepackage{wrapfig}

\usepackage[obeyFinal,colorinlistoftodos,textsize=scriptsize]{todonotes}
\usepackage{tikz}
\usetikzlibrary{fit,positioning,patterns,shapes,shapes.multipart,calc}
\usepackage{pgfplots}
\usepackage{pgfplotstable}
\pgfplotsset{compat=1.12}

\usepackage{amsthm}

\theoremstyle{definition}
\newtheorem{theorem}{Theorem}
\newtheorem*{theorem*}{Theorem}
\newtheorem{lemma}{Lemma}

\usepackage{lifty}

\newif\ifdraft
\drafttrue

\ifdraft
\newcommand\authorrnote[2]{\todo[color=#1]{#2}\xspace}
\newcommand{\TODO}[1]{{\color{orange!80!black}[\textsl{#1}]}}
\else
\newcommand\authorrnote[2]{}
\newcommand{\TODO}[1]{}
\fi

\newif\iflong
\longtrue

\newcommand{\toolname}{\textsc{Lifty}\xspace}
\newcommand{\fulltoolname}{Liquid Information Flow TYpes\xspace}
\newcommand{\corelg}{$\lambda^L$\xspace}

\newcommand{\Synquid}{\textsc{Synquid}\xspace}
\newcommand{\LH}{\textsc{LiquidHaskell}\xspace}
\newcommand{\LIO}{\textsc{LIO}\xspace}

\newcommand{\eg}{e.g.,\xspace}
\newcommand{\ie}{i.e.,\xspace}
\newcommand{\wrt}{\textit{wrt.}\xspace}

\newcommand{\custompar}[1]{\vspace{3pt}\noindent \textbf{\textit{#1}}~~}

\newcommand{\synquid}{\textsc{Synquid}\xspace}

\makeatletter

\newcommand\autorefs[1]{\@first@ref#1,@}
\def\@throw@dot#1.#2@{#1}
\def\@set@refname#1{
    \edef\@tmp{\getrefbykeydefault{#1}{anchor}{}}%
    \def\@refname{\@nameuse{\expandafter\@throw@dot\@tmp.@autorefsname}}%
}
\def\@first@ref#1,#2{%
  \ifx#2@\autoref{#1}\let\@nextref\@gobble
  \else%
    \@set@refname{#1}
    \@refname~\ref{#1}
    \let\@nextref\@next@ref
  \fi%
  \@nextref#2%
}
\def\@next@ref#1,#2{%
   \ifx#2@ and~\ref{#1}\let\@nextref\@gobble
   \else, \ref{#1}
   \fi%
   \@nextref#2%
}

\makeatother

\newcommand{\true}{\ensuremath{\mathsf{true}}}
\newcommand{\false}{\ensuremath{\mathsf{false}}}
\newcommand{\unit}{\ensuremath{\mathsf{()}}}
\newcommand{\tiocons}{\ensuremath{\mathsf{TIO}}\xspace}
\newcommand{\tioval}[1]{\ensuremath{\mathsf{TIO}~{#1}}}
\newcommand{\vtrue}{\ensuremath{\mathsf{True}}\xspace}
\newcommand{\vfalse}{\ensuremath{\mathsf{False}}\xspace}
\newcommand{\vuser}{\ensuremath{\mathsf{U}}\xspace}
\newcommand{\vfield}{\ensuremath{\mathsf{F}}\xspace}

\newcommand{\Flows}{\sqsubseteq}
\newcommand{\Fjoin}{\sqcup}
\newcommand{\Fmeet}{\sqcap}
\newcommand{\Implies}{\Rightarrow}

\newcommand\liff{\Leftrightarrow}

\newcommand{\Subt}{<:}

\newcommand\ontop[2]{\stackrel{\mathclap{\normalfont\scriptsize\mbox{#2}}}{#1}}

\newcommand{\env}{\ensuremath{\Gamma}\xspace}
\newcommand{\store}{\ensuremath{\Sigma}\xspace}
\newcommand{\funT}[3]{\ensuremath{{#2} \to {#3}}}

\newcommand{\config}[2]{\ensuremath{\langle{#1}\mid{#2}\rangle}}

\newcommand{\bigeval}{\ensuremath{\downarrow}\xspace}
\newcommand{\liobigeval}{\ensuremath{\Downarrow}\xspace}

\newcommand{\instr}[1]{\textcolor{red}{#1}}
\newcommand{\fact}[1]{(\textcolor{brown}{#1})}
\newcommand{\lequiv}[1]{\approx_{#1}}

\newcommand{\bvalues}{\ensuremath{\mathcal{B}}}
\newcommand{\fields}{\ensuremath{\mathcal{F}}}

\newcommand{\kw}[1]{\T{#1}\xspace}

\newcommand{\kwget}{\kw{get}}
\newcommand{\kwset}{\kw{set}}
\newcommand{\kwbind}{\kw{bind}}
\newcommand{\kwreturn}{\kw{return}}

\newcommand{\kwdown}{\kw{downgrade}}
\newcommand{\kwif}{\kw{if}}
\newcommand{\kwthen}{\kw{then}}
\newcommand{\kwelse}{\kw{else}}

\newcommand{\tio}[3]{\ensuremath{\T{TIO}\ {#1}\ \langle{#2, #3}\rangle}}
\newcommand{\ti}[2]{\ensuremath{\T{TI}\ {#1}\ \langle{#2}\rangle}}

\newcommand{\fld}[2]{\ensuremath{\T{Field}\ {#1}\ \langle{#2}\rangle}}
\newcommand{\cast}[2]{\langle{#2} \triangleleft {#1}\rangle}
\newcommand{\casts}{\hookrightarrow}
\newcommand{\typeof}{\mathsf{ty}}
\newcommand{\labelof}{\mathsf{lab}}
\newcommand{\erase}[2]{\ensuremath{\varepsilon_{#1}(#2)}}
\newcommand{\hole}{\ensuremath{\bullet}}
\newcommand{\sem}[1]{#1}

\newcommand*\Let[2]{\State #1 $\gets$ #2}
\algrenewcommand\algorithmicfunction{}

\makeatletter
\newcommand{\lookupGet}[1]{%
  \@ifundefined{lookup@#1}{%
    \PackageError{lookup}{No #1 key in lookup}{%
      Key #1 was not found in lookup. %
      You can insert it with \string\lookupPut{key}{value}.%
    }??%
  }{%
    \@nameuse{lookup@#1}%
  }%
}
\newcommand{\lookupPut}[2]{%
  \@namedef{lookup@#1}{#2}%
}
\makeatother

\begin{document}

\title{Liquid Information Flow Control}
\subtitle{Extended Version}

\author{Nadia Polikarpova}
\affiliation{
  \institution{University of California, San Diego}
}
\email{npolikarpova@eng.ucsd.edu}
\author{Deian Stefan}
\affiliation{
  \institution{University of California, San Diego}
}
\email{deian@cs.ucsd.edu}
\author{Jean Yang}
\affiliation{
  \institution{Carnegie Mellon University}
}
\email{jyang2@cs.cmu.edu}
\author{Shachar Itzhaky}
\affiliation{
  \institution{Technion}
}
\email{shachari@cs.technion.ac.il}
\author{Travis Hance}
\affiliation{
  \institution{Carnegie Mellon University}
}
\email{thance@cs.cmu.edu}
\author{Armando Solar-Lezama}
\affiliation{
  \institution{Massachusetts Institute of Technology}
}
\email{asolar@csail.mit.edu}

\renewcommand{\shortauthors}{N. Polikarpova et al.}

\begin{abstract}

We present \toolname, a domain-specific language for data-centric applications that manipulate sensitive data. 
A \toolname programmer annotates the sources of sensitive data with declarative security policies,
and the language statically and automatically verifies that the application handles the data according to the policies.
Moreover, if verification fails, \toolname suggests a provably correct repair,
thereby easing the programmer burden of implementing policy enforcing code throughout the application.

The main insight behind \toolname is to encode information flow control
using \emph{liquid types}, an expressive yet decidable type system.
Liquid types enable fully automatic checking of complex, data dependent policies,
and power our repair mechanism via type-driven error localization and patch synthesis.
Our experience using \toolname to implement three case studies from the literature shows that 
\begin{inparaenum}[(1)]
\item the \toolname policy language is sufficiently expressive to specify many real-world policies,
\item the \toolname type checker is able to verify secure programs and find leaks in insecure programs quickly, and 
\item even if the programmer leaves out \emph{all} policy enforcing code, 
the \toolname repair engine is able to patch all leaks automatically within a reasonable time. 
\end{inparaenum}


\end{abstract}

\maketitle

\section{Introduction}\label{sec:intro}

Modern applications handle sensitive user data in complex ways, subject to
increasingly complex \emph{security policies}.
For example, social networks like Twitter and Facebook must ensure that they
handle user data according to GDPR, health record systems (e.g., the Dexcom
diabetes management system) must abide by HIPAA, and financial applications
like Stripe and Mint must ensure they are PCI compliant.
In most cases, these applications even allow users to restrict who can access
their data---e.g., on Facebook a user can restrict access to (part of) their
profile to their friends.
Unfortunately, many applications specify and enforce these policies by strewing
checks throughout application code---an error prone process that has lead to
many inadvertent data leaks~\cite{pwn, privacyrights, unitedfail, walgreensfail}.

A promising approach to tackling this challenge is to use web frameworks like
Hails~\cite{GiffinLSTMMR12} and Jacqueline~\cite{Yang2016} which separate the security policy from the application
code and enforce the policy using dynamic information flow control
(IFC).
In such IFC frameworks, the programmer declaratively specifies expressive
data-dependent policies; the language runtime---or in the case of Hails, the
LIO monad~\cite{StefanM14,StefanMMR17}---then automatically enforces these policies to prevent
leaks (e.g., by throwing an exception or replacing sensitive values with
defaults).

Unfortunately, enforcing policies dynamically as in Hails and Jacqueline has
inherent limitations.
First, the IFC systems often perform redundant checks.
In many applications, developers already insert checks in the application code
to, for example, implement the user interface; alas, the IFC systems
do not know about these checks and will perform similar checks when enforcing
the policy.
These checks impose unnecessary performance overheads, i.e., they tax the
application latency a second time.
Second, errors due to policy violations only manifest at runtime: the
programmer doesn't know if their policy is too strict until their application
crashes at runtime.

Static IFC systems
(e.g.,~\cite{Jif0,Li2005,Zheng2007,Russo08,JiaZ09,SwamyCC10,Chlipala10,buiras2015hlio})
precisely address these limitations: they do not impose unnecessary runtime
checks and catch errors early---at compile time.
Unfortunately existing static IFC systems either lack support for expressive
data-dependent policies (necessary in modern applications), or they require
manual proofs or annotations to be strewed thought the application code.
 
In this paper, we take the best from both worlds: we present a static IFC system that
automatically---without manual proofs or annotations---enforces Hails-like
expressive, declarative policies.

\custompar{The \toolname language.}
Our first insight is that we can encode static IFC into the framework of \emph{liquid types}~\cite{RondonKaJh08,VazouRoJh13,VazouSeJhVyJo14},
an expressive yet decidable type system.
To this end, we use predicates from decidable logics to directly specify
expressive policies and adopt security monads~\cite{LiZ06,Russo08,VassenaR16} to enforce these
policies.
We do this in \toolname---short for \fulltoolname---a domain-specific
language (DSL) for writing secure data-centric applications.
With \toolname, programmers (1) write code in our custom \T{IO} monad called
\T{TIO} and (2) specify policies in a decidable logic when declaring sources of
sensitive data.
The \toolname type-checker uses liquid types to verify the program against the
policies and flags any \emph{unsafe access} to sensitive data as a type error.

\begin{figure}
\begin{minipage}{0.49\textwidth}
\centering
  \includegraphics[width=.9\columnwidth]{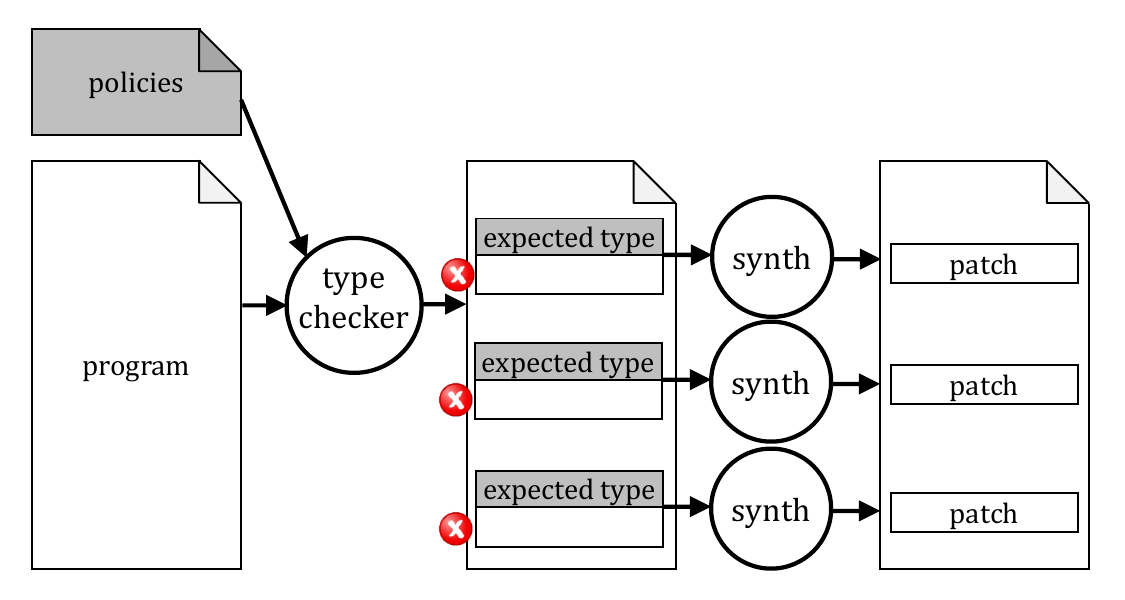}
  \caption{Patching information leaks with \toolname.}\label{fig:overview_diagram}
\end{minipage}
\hfill
\begin{minipage}{0.49\textwidth}
\centering
  \includegraphics[width=.9\columnwidth]{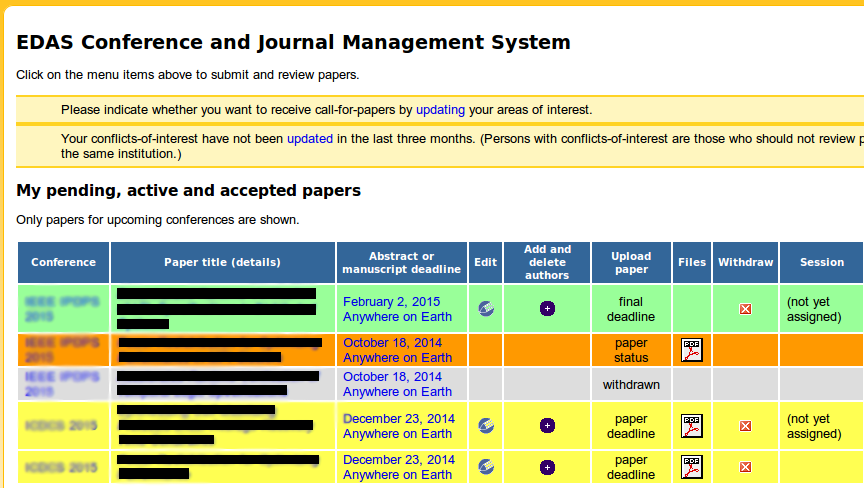}
  \caption{Author's home screen in EDAS, shared with permission of Agrawal and Bonakdarpour~\cite{Agrawal16}.}
  \label{fig:EDAS_leak}
\end{minipage}
\end{figure}

\custompar{Leak Repair.}
By taking a static approach to enforcing information flow, \toolname can also
help programmers repair their unsafe code.
Our insight here is to use type errors to localize the source of each leak and
suggest a best-effort \emph{leak patch} (\autoref{fig:overview_diagram}).
The suggested patches guard the unsafe access with a policy check and, for the
failure case, implement a safe escape---e.g., they return a default value.
%
The key to the efficiency of our repair technique
is a new \emph{leak localization} mechanism
that relies on the \toolname type-checker to infer an expected type
for each unsafe access.
While efficient, this approach is necessarily limited: 
although the patch is guaranteed to fix the leak,
the generated policy check might be conservative,
or the repair attempt might fail (\eg when it cannot find a safe escape).
Nevertheless, we find this best-effort useful in practice.

\custompar{Evaluation.}
We evaluated our prototype implementation of \toolname on a series of small, but representative micro-benchmarks,
as well as three case studies: a conference manager, a health portal, and a student grade record system.
For each of these programs, we write the code omitting all policy checks (\ie as if all the data were publicly visible),
and ask \toolname to repair the code---and make it secure.
Our evaluation demonstrates that our solution supports expressive policies
and is able to generate all necessary patches for our benchmarks
within a reasonable time (within a minute for each of our case studies).
%

\custompar{Contributions.}
In summary, we make the following contributions:
\begin{enumerate}
\item The \T{TIO} Monad: an encoding of static IFC into liquid types, that supports fully automatic verification of expressive policies.
\item Leak repair: we combine novel leak localization with type-driven synthesis to generate best-effort patches for the leaks.
\item Prototype implementation: we implement a prototype of \toolname 
  and evaluate our system on several micro-benchmarks and case studies.
\end{enumerate}

\section{Motivating Example}\label{sec:example}
We motivate \toolname using an example based on a leak from the EDAS conference manager~\cite{Agrawal16}.
In this section, we describe the anatomy of this leak,
show how \toolname can detect it at compile-time given a declarative security policy, 
and demonstrate how our tool automatically synthesizes a patch for this leak.
The core technical innovation that enables automatic verification and synthesis---the \toolname type system---%
is introduced in \autoref{sec:overview},
together with more advanced examples that demonstrate the flexibility of our language.

\subsection{The EDAS Leak}


\autoref{fig:EDAS_leak} shows a screenshot of the EDAS conference manager home
screen.
On the home screen, users are presented with an overview of all their submitted
papers (both old and new).
Color coding indicates PC decisions: green papers have been accepted, orange
have been rejected, and yellow papers are awaiting notification. 
As usual, users are not supposed to learn the acceptance decision of their
papers before the notifications are out.
But, the site is leaky: in the figure, we can infer that the first one of the
yellow papers has been tentatively accepted, while the second one has been
tentatively rejected.
We can make this conclusion because the two rows differ in the value of the
``Session'' column---and sessions are only displayed for accepted papers.

This leak is particularly insidious---indeed, it's an example of an
\emph{implicit flow}: the ``accepted'' decision does not appear anywhere on the
screen, but it does conditionally influence the output of the web applicatin.
To prevent such leaks, it is insufficient to simply examine output values; we
must track the flow of sensitive information throughout the system.

\begin{figure}
\small
\begin{minipage}[t]{0.5\textwidth}
\begin{lifty}[numberblanklines=false,showlines=true]
showPaper ds client p = do 
  t   <- getTitle ds p
  dec <- getDecision ds p
  if dec == Accept 
    then do
      ses <- getSession ds p
      print client (t + " " + ses)
    else print client t    
\end{lifty}
\vspace{1.28cm}
\caption{\label{code:edas-leak} Simplified version of the EDAS leak.}
\end{minipage}%
\begin{minipage}[t]{0.5\textwidth}
\begin{lifty}
@showPaper ds client p = do 
  t   <- getTitle ds p
  dec <- @do x1 <- getPhase ds
            if x1 == Done 
              then @getDecision ds p@ 
              else return NoDecision@
  if dec == Accept 
    then do
      ses <- getSession ds p
      print client (t + " " + ses)
    else print client t@    
\end{lifty}
\caption{\label{code:edas-fixed} With a leak patch inserted by \toolname.}
\end{minipage}
\end{figure}

\autoref{code:edas-leak} shows a simplified version of the leaky code (in
\toolname syntax) that is used to display the description of individual papers.
This code retrieves the title and decision for paper \T{p}
and, if the paper has been accepted,
it retrieves the session where the paper will be presented and displays it to
\T{client} together with the title;
otherwise, it only displays the title%
\footnote{The \T{ds} parameter models the state of the \emph{data store};
it is only used for specification purposes and is threaded through \toolname programs explicitly for simplicity.}.
In deciding to display the session (or not), this code indirectly leaks the
decision to \T{client}.
 
The easiest way to fix this leak is to check if the conference is in an
appropriate phase (reviewing is done), and only then display the session.
But, even for a simple example like EDAS, we must strew in such
\emph{policy-enforcing code} in the dozen of web request handlers that access
papers, reviews, etc.
Real web applications handle lots of sensitive data and have hundreads to
thousands of such handlers; getting all the right checks in all the right
places is notoriously hard.


\subsection{Programming with \toolname}

In \toolname, the programmer explicitly associates sensitive data with declarative policies
by annotating \emph{input actions} that retrieve data from the store with appropriate types.
For example, to specify that PC decisions are visible to everyone once the
conference phase is \T{Done} but not otherwise,
the action \T{getDecision} in the EDAS example can be annotated with the type: 
{\small
\begin{lifty}[numbers=none]
getDecision :: ds:Store -> p:PaperId -> TIO Decision <phase ds = Done, false>
\end{lifty}
}
The \T{TIO} type constructor is indexed by two security \emph{labels}:
the first label specifies when a user is allowed to see
the result of this action;
we postpone the explanation of the second label to \autoref{sec:overview}.
Labels can be expressive, value-dependent predicates as in this example---here,
the label depends on the state of the data store \T{ds}---or simple predicates
(e.g., input actions for public fields ``title'' and ``session'' are labeled
\T{true}).

Given such policy-annotated actions, \toolname would statically reject the code
in \autoref{code:edas-leak} with a \emph{type error}: the value of \T{dec}
obtained on line 3 is flowing to \T{client}, but is not visible to \T{client}
in the current state \T{ds}.
Moreover, \toolname would suggest a \emph{patch} for this leak, as shown in
\autoref{code:edas-fixed}.
The fixed code guards the access to the sensitive field ``decision'' with a
policy check, and if the check fails, it substitutes the true value of
this field with a default value---a constant \T{NoDecison}.
\toolname, in turn, guarantees that the patched code respects the declared
policies.

To our knowledge, only two other information-flow control tools support
similarly expressive declarative policies.
The first is LIO, as used in the Hails web framework~\cite{GiffinLSTMMR12,
giffin:2017:hails}, or the LWeb framework~\cite{parker2019lweb}.
LIO does not check policies statically.
Hence, the equivalent code of \autoref{code:edas-leak} would compile without
errors.
Instead, at run time, LIO would throw an exception on either line 7 or 8,
when trying to execute the \T{print} action (after inspecting the decision).
Though this successfully prevents the leak, it also means the programmer won't
find out that their application is broken until run time.

The second tool is Jeeves as used in Jaqueline framework~\cite{Yang2016}.
As with LIO, the equivalent Jeeves code of \autoref{code:edas-leak} would compile
without errors.
The runtime behavior of the Jeeves code, however, is identical to the version
patched by \toolname, \ie the value of \T{dec} will be replaced by a default
whenever the conference is not \T{Done}. 
Unlike \toolname though, Jeeves achieves this using \emph{faceted
execution}~\cite{Yang2012}, i.e., by performing the computation on multiple
versions of each sensitive value, which can have prohibitive runtime
overhead.
Moreover, Jeeves replaces sensitive values with default values implicitly, at
run time---this makes it harder for the programmer to modify the tool's default
mitigation strategy and find bugs due to default values.

\section{Overview}\label{sec:overview}

In this section we first describe how \toolname uses liquid types to enforce
static information flow control.
We then explain how to use this IFC mechanism to implement secure data-centric applications
with complex, data-dependent policies.
Next, we give the intuition for how \toolname's type-driven repair engine
generates leak patches for these application.
%
We wrap up the section with more advanced examples from the conference manager.
%


\subsection{Static IFC with \T{TIO}}\label{sec:overview:tio}

\custompar{Liquid types.}
\toolname builds upon a pure functional language with \emph{liquid types}~\cite{RondonKaJh08,VazouRoJh13}---%
types decorated with predicates from SMT-decidable logics.
For example, we can define the type of natural numbers as \T{type Nat = \{Int|0 <= _v\}},
where $\nu$ is a reserved \emph{value variable}, which ranges over the values of the refined type.
State-of-the-art liquid type systems feature subtyping (\eg $\T{Nat} \Subt \T{Int}$),
as well as type constructors that can be indexed by both types and refinement predicates.
For example, we can define a type constructor \T{List _a <p: _a -> _a -> Bool>}
for lists with elements of type $\alpha$,
where each in-order pair of elements satisfies a binary predicate $p$;
then the type \T{List Nat <\\x y.x <= y>} denotes a sorted list of natural numbers.
Concretely our implementation builds upon the \synquid language~\cite{PolikarpovaKuSo16},
but a similar technique could be used to add static IFC to \LH~\cite{VazouSeJh14}.

\custompar{Security lattice.}
Like almost all IFC systems, \toolname uses a security lattice to distinguish
between data with different levels of confidentiality.
We, however, fix the security lattice to be the lattice of
\emph{refinement-logic predicates over principals}.
More precisely, a \emph{security label} $\ell: \T{User}\to \T{Bool}$
denotes information visible to users that satisfy $\ell$%
\footnote{For simplicity, we use a concrete type \T{User} to denote principals; 
our approach also supports leaving this type as a parameter.}.
For example, a key shared between \T{alice} and \T{bob} has a security label $\lambda u.u \in [\T{alice},\T{bob}]$.
In the rest of the paper we will use a reserved variable $\upsilon$ to range over users in security labels
and omit the $\lambda$-binding;
for example, the label introduced above is written as $\upsilon \in [\T{alice},\T{bob}]$.
Note that the usual ``can-flow-to'' partial order in our security lattice
corresponds to reverse implication: $l \Flows h$ iff $\forall \upsilon . h \Implies l$, 
and hence the bottom (least secret) label is $\bot = \true$ and the top
(most secret) label is $\top = \false$.
Our lattice can be seen as a generalization of DCLabels~\cite{dclabels} and
is thus at least as expressive~\cite{GenLabels}.
%

\custompar{The \T{TIO} monad.}
To extend the base language with static IFC, we follow a long line of work on
\emph{security monads}~\cite{LiZ06,Russo08,Russo15,VassenaR16}---in particular
SLIO~\cite{buiras2015hlio}---where sensitive computations are wrapped in a
special datatype indexed by a security label. 
Proper assignment of labels and their propagation through the program
are ensured by the types of primitives in the API of the security monad.

More specifically, \toolname introduces the type constructor \T{TIO} (``tagged input-output''),
indexed by a return type and two security labels, 
the \emph{input label} $i$ and the \emph{output label} $o$.
The type \tio{T}{i}{o} denotes a secure computation
that may read from resources at security level $i$ (and below)
and may write to resources at security level $o$ (and above).
For example, a computation \T{getSharedKey} that reads a key shared between \T{alice} and \T{bob} can have the type 
\tio{\T{String}}{\upsilon \in [\T{alice},\T{bob}]}{\false}
to indicate that its result should be viewed only by \T{alice} and \T{bob},
and that it does not have any (user-visible) output effects.

\custompar{Subtyping.}
%
The \T{TIO} constructor is \emph{covariant} in the input label
and \emph{contravariant} in the output label \wrt the lattice ordering,
\ie $\tio{T}{i_1}{o_1} \Subt \tio{T}{i_2}{o_2}$ iff $i_1 \Flows i_2$ and $o_2 \Flows o_1$.
For example, we can pass the \T{getSharedKey} computation from above
to a function \T{f} with argument $x\colon \tio{\T{String}}{\upsilon = \T{alice}}{\upsilon = \T{alice}}$.
Intuitively this is safe, because \T{f} requires the result of $x$ to be visible at least to \T{alice}
(while in fact the shared key is visible to both \T{alice} and \T{bob}),
and counts on $x$ to output at most to \T{alice}
(while in fact \T{getSharedKey} does not perform any user-visible output).
Because lattice ordering reduces to implication between refinement formulas,
the subtyping between \T{TIO} types can be automatically decided by the base language type checker
(with the help of an SMT solver).


\begin{figure}
\small
\begin{minipage}[t]{.5\textwidth}
\begin{lifty}[numbers=none]
getSharedKey :: TI String <_u _in_ [alice,bob]>
getSSN :: x:User -> TI String <_u = x>
\end{lifty}
\end{minipage}%
\begin{minipage}[t]{.5\textwidth}
\begin{lifty}[numbers=none]
print :: x:User -> String -> TO () <_u = x>
\end{lifty}
\end{minipage}
\caption{\label{code:io_actions} Example input/output actions.}
\end{figure}

\custompar{Input/output actions.}
\toolname programmers identify sources and sinks of sensitive information
by providing a set of domain-specific atomic \emph{input} and \emph{output actions},
annotated with labels that are assumed to be correct.
\autoref{code:io_actions} gives an example set of atomic actions
that includes reading the shared key of users \T{alice} and \T{bob}, 
reading a user's social-security number (which is visible only to the user themselves),
and printing a string to a given user.
\T{TI} and \T{TO} are type synonyms for \T{TIO} computations that perform only input or only output, respectively
(see \autoref{code:primitives}).

Similar to other IFC systems (e.g., LIO and SLIO), we leave \toolname agnostic
to the particular choice of input/output actions; depending on the domain,
actions can be used to model reading and writing to mutable memory, file
system, or database, as well as HTTP responses, sending emails, etc.

\begin{figure}
\small
\begin{minipage}{0.5\textwidth}
\begin{lifty}
-- Primitives (trusted)
data TIO _a <i,o> = -- hidden from clients
return :: /\_a._a -> TIO _a <true, false>
bind :: /\_a,_b,i,j,o,p.TIO _a <i, o> 
 -> (_a -> TIO _b <j, i && p>) 
 -> TIO _b <i && j, o || (p && i)>
downgrade :: /\ c,i,o.
  TIO {Bool|_v => c} <i && c> <o> 
  -> TIO {Bool|_v => c} <i> <o>
\end{lifty}
\end{minipage}\hfill
\begin{minipage}{0.5\textwidth}
\begin{lifty}[firstnumber=10]
-- Auxiliary (typed-checked)
type TI _a <i> = TIO _a <i, False>
type TO _a <o> = TIO _a <True, o>
seq :: /\_a,_b,i,j,o,p.TIO _a <i,o>
 -> TIO _b <j,p> -> TIO _b <j,o || p>
mapM :: /\_a,_b,i. (_a -> TIO _b <i,i>) 
  -> [_a] -> TIO [_b] <i,i>
... -- liftM, filterM, sortByM
\end{lifty}

               
\end{minipage}
\caption{\label{code:primitives} Excerpt from the \T{TIO} API.}
\end{figure}

\custompar{\T{TIO} Primitives.}
Client code manipulates \T{TIO} computations through the API shown in \autoref{code:primitives}.
The API consists of three core primitives;
these form the trusted computing base of \toolname
(together with the language runtime and type checker).
%
The API also exposes several auxiliary functions
that are built on top of the primitives and verified using the type checker.

The first core primitive, \T{return}, simply embeds a pure value into a sensitive computation,
and hence has the strongest possible labels: input label $\bot$ and output label $\top$.
The \T{bind} primitive (the analogue of \lstinline[language=Haskell]^>>=^ in Haskell) sequences two sensitive computations, $a$ and $b$,
such that $a$'s result flows into $b$.
To prevent leaks, we need to guarantee that $a$'s input label can flow to $b$'s output label.
To enforce this, we simply add $a$'s input label \T{i} as a \emph{conjunct} to $b$'s output label.
Moreover, we set the input label of $\T{bind}\ a\ b$ to the join (\emph{conjunction}) of the input labels of $a$ and $b$,
and its output label to the meet (\emph{disjunction}) of the output labels of the two.
Finally, the \T{downgrade} primitive supports safe declassification of boolean terms;
we postpone its presentation to the end of this section.

\custompar{Auxiliary \T{TIO} API functions.}
Often a \toolname program sequences two computations $a$ and $b$,
but no information is actually flowing from $a$ to $b$.
To relax the restrictions imposed by \kwbind in this case,
we provide an API function \T{seq} (the analogue of \lstinline[language=Haskell]^>>^ in Haskell).
The type of \T{seq} does not enforce any relationship between the labels of $a$
and $b$.
Note that \T{seq} is not a primitive, 
and can be implemented using \T{bind} and \T{downgrade}.

The rest of the \T{TIO} API contains monadic combinators such as
\T{liftM} (for lifting a pure function into \T{TIO}),
\T{mapM} (for mapping a sensitive computation over a list)%
\footnote{%
Why are the input and output labels of \T{mapM} the same?
The only way to sequence two \T{TIO} computations and preserve both of their return values is to use \kwbind. 
But \kwbind requires that the input label of the first computation can flow to the output label of the second; 
hence \T{mapM} only verifies if its input label can flow to its output label.
This is, of course, conservative: the later iterations of \T{mapM} do not actually use the results of the previous iterations! 
To express this fact, we could equip \T{TIO} with the \emph{applicative} interface in addition to the \emph{monadic} one,
introducing a primitive operation for running two \T{TIO} computations independently and then combining their results. 
We decided not to pursue this path in the interest of simplicity.%
},
as well as \T{filterM} and \T{sortByM} (for filtering and sorting based on sensitive criteria).
All auxiliary functions are implemented in terms of the three primitives 
and type-checked automatically by \toolname in less than a second.

\begin{figure}
\small
\begin{minipage}{0.24\textwidth}
\begin{lifty}
ok1 = 
 do
  k <- getSharedKey
  print bob k
  print alice k  
\end{lifty}
\end{minipage}\hfill
\begin{minipage}{0.23\textwidth}
\begin{lifty}
bad1 = 
 do 
  b <- getSSN bob
  print bob b
  print alice b
\end{lifty}
\end{minipage}\hfill
\begin{minipage}{0.26\textwidth}
\begin{lifty}
bad2 = 
 do b <- getSSN  bob
    print bob b
    a <- getSSN alice
    print alice a
\end{lifty}
\end{minipage}\hfill
\begin{minipage}{0.27\textwidth}
\begin{lifty}
ok2 = 
 do do b <- getSSN bob
       print bob b
    a <- getSSN alice
    print alice a
\end{lifty}
\end{minipage}
\caption{\label{code:toy-ex} Examples of well-typed (\T{ok1, ok2}) and ill-typed (\T{bad1, bad2}) \T{TIO} computations.}
\end{figure}

\custompar{Examples.}
To gain some intuition about our IFC encoding, consider simple examples of \T{TIO} computations in \autoref{code:toy-ex}.
\toolname's surface syntax supports Haskell-like \T{do}-notation~\cite{Haskell2010},
which desugars into invocations of \T{bind} and \T{seq} in a standard way.
For example, \T{ok1} is desugared into \T{bind getSharedKey (\\k.seq (print bob k) (print alice k))}.
Note that the desugaring uses \T{seq} (with its more permissive type) instead of \T{bind}
whenever the return value of a line is not bound.

These snippets use atomic actions defined in \autoref{code:io_actions}.
The snippet \T{ok1} is well-typed in \toolname
because the output label of the sequence of two \T{print} actions, $\upsilon = \T{bob} \vee \upsilon = \T{alice}$,
implies the input label of \T{getSharedKey}, $\upsilon \in [\T{alice},\T{bob}]$.
More precisely, we can instantiate the type of \T{bind} with $\T{i} \mapsto \upsilon \in [\T{alice},\T{bob}]$, $\T{p}, \T{j}\mapsto \true$, $\T{o}\mapsto \false$.
On the other hand, the snippet \T{bad1} is ill-typed because the implication no
longer holds---indeed, this would be leaking \T{bob}'s SSN to \T{alice}.

Perhaps surprisingly, the snippet \T{bad2} is also ill-typed:
we cannot bind \T{getSSN bob} to the rest of the computation,
whose output label permits output to \T{alice}.
The code does not actually leak \T{b} to \T{alice}.
Instead, \toolname flags this snippet because our information flow tracking is
\emph{coarse-grained}~\cite{buiras2015hlio,StefanMMR17,VassenaR16},
and restricts outputs based on all the data \emph{in scope};
in particular, the output to \T{alice} on line 5 should be rejected
because \T{b}, which is not visible to \T{alice}, is still in scope on line 5.

This coarse-grained approach to IFC is simpler, but as expressive as
fine-grained IFC~\cite{rajani2019expressiveness}.
We simply need to tweak the code to express the desired program:
\T{ok2} performs the same actions in the same order,
but with a different binding structure.
Here lines 2 and 3 handling \T{bob}'s data are grouped into one \T{TIO} computation,
while lines 4 and 5 handling \T{alice}'s data are grouped into another \T{TIO} computation;
the two actions are then sequenced with \T{seq}, 
which lets the type checker know that no information flows between the two,
leading to a well-typed program.

\custompar{Type checking}
The \toolname type checker is based on the \emph{liquid types} inference framework~\cite{RondonKaJh08,CosmanJ17}.
To type-check a \T{TIO} computation, it uses the types of the API functions
to generate a system of subtyping constraints over \T{TIO} types.
It then relies on the definition of co- and contravariant subtyping
to reduce subtyping constraints to a system of \emph{constrained Horn clauses} (CHCs),
\ie implications between (possibly unknown) refinement predicates.
For example, the snippet \T{ok1} from \autoref{code:toy-ex}
generates the following CHCs (trivial constraints are omitted):
\begin{align}
&I \wedge \neg(\upsilon \in [\T{alice}, \T{bob}]) \Implies \false \label{eq:toy-query}\\
&(O \vee P) \Implies I \label{eq:toy-rule-one}\\
&\upsilon = \T{alice} \Implies P  \label{eq:toy-rule-two}\\
&\upsilon = \T{bob} \Implies O  \label{eq:toy-rule-three}
\end{align}
Here $I, O, P$ are unknown predicates over the reserved variable $\upsilon$
and any program variables in scope;
these unknowns stand, respectively, for the instantiations of indexes \T{i} in \T{bind} on line 3 
and \T{o} and \T{p} in \T{seq} on line 4.
Specifically, constraint (1) relates \T{bind} to \T{getSharedKey},
constraint (2) relates the output label of the \T{seq} computation to the input label of the left-hand-side of \T{bind},
while the last two constraints relate the output labels of \T{print} actions to the output label of \T{seq}.
The left- and right-hand sides of the implication are called the \emph{body} and the \emph{head} of a CHC, respectively.
CHCs can be divided into \emph{rules} (head is an unknown, like (\ref{eq:toy-rule-one})--(\ref{eq:toy-rule-three})) 
and \emph{queries} (head is $ \false$, like (\ref{eq:toy-query})).
In this case, the CHCs are non-recursive, and hence can be solved by simple left-to-write unfolding of the rules:
from (\ref{eq:toy-rule-two}) and (\ref{eq:toy-rule-three}) we infer the \emph{strongest} assignment to $O$ and $P$: 
$[O\mapsto \upsilon = \T{bob},  P\mapsto \upsilon = \T{alice}]$;
substituting this into (\ref{eq:toy-rule-one}), we similarly infer $[I \mapsto (\upsilon = \T{alice} \vee \upsilon = \T{bob})]$;
finally, with this assignment the query (\ref{eq:toy-query}) is valid,
hence this assignment is a (strongest) solution. 
Recursive CHCs are a bit more involved,
but we can still find the strongest solution using a combination of unfolding and predicate abstraction, as shown in~\cite{CosmanJ17}. 

\custompar{Safe downgrading.}
%
We leverage the power of refinement types to support safe, \ie non-leaky,
declassification of boolean terms.
We do this with the \T{downgrade} primitive.
The intuition behind \T{downgrade t}
is that the input label of \T{t} can be safely lowered
as long as we can prove that in all relevant executions \T{t} always returns the same value 
(in particular, the value \T{False}),
because constants cannot leak information.
Consider the following term:
\begin{lifty}[numbers=none]
downgrade (bind (getSSN x) (\s.return valid(s) && x == alice))
\end{lifty}
In \toolname, this term type-checks against the type \ti{\T{Bool}}{\T{_u = alice}}
even though it performs an input action of an incompatible type \ti{\T{Bool}}{\T{_u = x}}.
Intuitively, this is safe because in all executions where the two input labels are indeed incompatible,
it must be that $\T{x} \neq \T{alice}$,
and hence the \emph{result} of the computation is always \T{False}.
\toolname performs this reasoning automatically by instantiating the type of \kwdown shown in \autoref{code:primitives} with
$i \mapsto \T{_u = alice}, c \mapsto \T{x = alice}$.
Note that this type would no longer work if we removed the conjunct \T{x == alice} from the term.

Safe downgrading in \toolname is restricted to boolean terms,
which lets us rely on existing machinery of liquid type inference 
to discover all intermediate labels completely automatically.
Although such a restrictive mechanism might not appear very useful at first,
it turns out to be indispensable for supporting applications with complex data-dependent policies,
as we demonstrate in \autoref{overview:advanced}.
Finally, as we mentioned above, \kwdown can be used to implement \T{seq}:
\begin{lifty}[numbers=none]
seq a b = bind (downgrade (bind a (\_.return False))) (\_.b)
\end{lifty}

\subsection{Encoding Policies in Data-Centric Applications}\label{sec:overview:policies}

The \T{TIO} monad is particularly suitable for enforcing secure information
flow in data-centric web applications (such as a conference manager or a social
network).
Such applications are built around a \emph{data store}, 
where different fields have different visibility policies,
which might depend on the data itself. 
Web application are typically structured as a set of controllers or request
handlers, \ie functions called by a user request that read data from the store,
process it and then respond to the user.
A \toolname programmer can encode store reads and writes as atomic input and output actions,
and responses as output actions.
They can directly express data-dependent policies as types, instead of
translating them into an intermediate security lattice.

\begin{figure}
\small
\begin{lifty}
-- Datatypes and redaction functions
data Decision = Accept|Reject|NoDecision
redact NoDecision

-- Specification-only function for each field:
measure phase   :: Store -> Phase
measure authors :: Store -> PaperId -> Set User

-- Field getters as TIO input actions:
getPhase    :: ds:Store              -> TI {Phase|_v = phase ds}            <True>
getTitle    :: ds:Store -> p:PaperId -> TI String                           <True>
getDecision :: ds:Store -> p:PaperId -> TI Decision                         <phase ds == Done>
getAuthors  :: ds:Store -> p:PaperId -> TI {[User]|elems _v = authors ds p} <_u _in_ authors ds p 
                                                                           || phase ds == Done>
-- Field setters as TIO output actions:
setDecision :: ds:Store -> dec:Decision -> TO Store                         <phase ds == Done>
\end{lifty}
\caption{\label{code:edas-policies} Excerpt from the data store API for the conference manager.}
\end{figure}

\custompar{Conference manager.}
Let us revisit the conference manager example from \autoref{sec:example}
and demonstrate how a \toolname programmer would specify its data-dependent policies.
\autoref{code:edas-policies} shows an excerpt from the data store API for this system.
To encode the policies, we introduce an uninterpreted type \T{Store},
which models the state of the data store.
Next, for each field of the store we introduce a \emph{measure},
\ie an uninterpreted function that models the value of the field.
Note that both of these are specification-only constructs, 
introduced solely for the purpose of expressing policies.

The actual program-level API of the data store contains a \emph{getter} and a \emph{setter}
for each field of the store, encoded as atomic \T{TIO} actions.
The programmer, however, can use the uninterpreted measures to relate the return values of the getters
to the labels of the actions.
For example, in \autoref{code:edas-policies},
we use the measure \T{phase} both in the return type of \T{getPhase} 
and in the input label of \T{getDecision} (resp. output label of \T{setDecision}),
thereby relating the two actions.

The reason \toolname considers the EDAS leak example from
\autoref{code:edas-leak} ill-typed is now clear:
the input action \T{getDecision ds p} on line 3 has the input label $\T{phase ds} = \T{Done}$,
but this action is bound to a computation with the output label $\upsilon = \T{client}$.
For this occurrence of \T{bind} to be well-typed we need to prove that $\upsilon = \T{client} \Implies \T{phase ds} = \T{Done}$ is valid,
which does not hold.

On the other hand, the patched code in \autoref{code:edas-fixed} is well-typed.
To understand why, note that due to the polymorphic type of \T{bind},
the type of the binder \T{x1} on line 3 is \T{\{Phase|_v = phase ds\}}.
Hence the input action \T{getDecision ds p} is now being type-checked \emph{under the assumption} $\T{Done} = \T{phase ds}$,
which makes the above implication valid.

\subsection{Patching the Leaks}

How does \toolname patch the leak in \autoref{code:edas-leak}?
Intuitively, our goal is to eliminate the type error in this program
by breaking the insecure flow from the input action \T{getDecision ds p} on line 3
into the output actions \T{print client} on lines 7 and 8.
There are, of course, many ways to break this flow,
and in the absence of a complete functional specification for \T{showPaper},
\toolname cannot be sure what the programmer indented.
However, in the domain of data-centric applications there is a reasonable default:
\toolname can \emph{guard} the offending input action with a policy check,
and \emph{redact} the sensitive value whenever the policy check fails.
%
\toolname borrows this repair strategy from Jeeves~\cite{Yang2012,Yang2016}, 
which enforces policies by replacing secret values with public defaults.
Note, however, that a \toolname programmer does not have to use this strategy:
because repair happens statically,
they can inspect the suggested patch and if necessary replace it with a manual fix.

\toolname implements this leak repair strategy in three steps:
\begin{inparaenum}
\item localize leaky input actions,
\item for each such input action, infer the \emph{expected type} of the patch (\ie the weakest type that would eliminate the type error),
and \item generate a term of the expected type by filling a domain-specific template.
\end{inparaenum}
We describe these steps in the rest of this section.

\custompar{Localizing leaks.}
Type-checking the code in \autoref{code:edas-leak} 
against the policies in \autoref{code:edas-policies} generates the following (simplified) system of CHCs:
\begin{align}
&I \wedge \T{phase ds} = \T{Done} \Implies \false \label{eq:query}\\
&O \Implies I \label{eq:rule1}\\
&\upsilon = \T{client} \Implies O \label{eq:rule2}
\end{align}
This system clearly has no solution;
in particular, unfolding the rules (\ref{eq:rule1})--(\ref{eq:rule2}) gives the strongest assignment $I\mapsto \upsilon = \T{client}$,
but this assignment does not validate the query (\ref{eq:query}).
Our insight is that each such invalid query corresponds to an atomic input action 
whose input label is too high for the output action it is flowing to.
Using this insight, the \toolname type checker can identify all leaky input actions at the same time,
with just a little extra bookkeeping---namely, tracking which term generated which CHC.
In this example, the query (\ref{eq:query}) is generated by \T{getDecision ds p},
so this is the action we need to guard.

\custompar{Inferring the expected type.}
From the same CHCs 
we can infer not only the offending term,
but also the highest input label a replacement term can have for the program to type-check.
We obtain this label from the strongest assignment computed from the rule clauses.
In our example, the strongest assignment has $I\mapsto \upsilon = \T{client}$,
hence replacing \T{getDecision ds p} with \emph{any term} of type \ti{\T{Decision}}{\T{_u = client}}
would fix the leak.
We refer to this type as the \emph{expected type} of the patch.

\custompar{Synthesizing the patch.}
%
%
Even though \emph{any} term of the expected type is secure, 
not all solutions are equally desirable:
for example, we wouldn't want the patch to return \T{Accept} unconditionally.
Intuitively, a desirable solution returns the original value whenever it is safe,
and otherwise replaces it with a reasonable redacted value.
\toolname achieves this through a combination of two measures.
First, instead of synthesizing a single term of type \ti{\T{Decision}}{\T{_u = client}},
it generates a set of candidate \emph{branches}
(by enumerating all branch-free terms of this type up to a fixed size).
Second, \toolname gives the programmer control over the space of possible redacted values
by generating the branches in a restricted environment,
which only contains the original term and explicitly designated \emph{redaction functions}.
When defining a new datatype, the programmer is expected to designate one or more constructors (or functions)
of this type as redactions
(\autoref{code:edas-policies} shows an example for type \T{Decision}).
As a result, our running example generates only two branches:
\begin{align*}
&\T{getDecision ds p} :: \ti{\T{Decision}}{\T{phase ds} = \T{Done}}\\
&\T{return NoDecision} :: \ti{\T{Decision}}{\true}
\end{align*}

Next, for every branch, 
\toolname attempts to abduce a condition that would make the branch type-check against the expected type.
In our example, the second branch is correct unconditionally, 
while the first branch generates the following \emph{logical abduction} problem:
$
C \wedge \upsilon = \T{client} \Implies \T{phase ds} = \T{Done},
$
where $C$ is an unknown formula over only the program variables,
\ie it cannot mention the user variable $\upsilon$.
\toolname uses existing techniques~\cite{PolikarpovaKuSo16} to find the following solution
to the abduction problem: 
$C\mapsto \T{phase ds} = \T{Done}.$
It then sorts all successfully abduced branch conditions from strongest to weakest,
and uses each condition to synthesize a \emph{guard},
\ie a program that computes the monadic version of the condition.
In our case, the guard for the first branch is
\T{bind getPhase (\\x1.x1 = Done)}.
Finally, \toolname combines the synthesized guards and branches into a single conditional,
which becomes the patch and replaces the original offending input action.

\subsection{Advanced Policies}\label{overview:advanced}

We conclude the overview of \toolname with another example from the conference
manager, which illustrates the kinds of policies we need to realistically express.

\custompar{Self-referential policies.}
Consider the action \T{getAuthors} in \autoref{code:edas-policies}
that retrieves the author list of a given submission.
Assuming that our conference is double-blind,
we would like to enforce a policy that the author list is only visible to the authors themselves until the reviewing is done
(and afterwards visible to everyone).
Unlike the policy on field ``decision'', 
which depends on a \emph{public} field ``phase'',
this is an example of a policy that depends on a \emph{sensitive} field;
moreover, in this case the policy is \emph{self-referential}:
it guards access to ``authors'' in a way that depends on the value of ``authors''.
By separating measures from input/output actions, \toolname makes
it easy to express such self-referential policies:
the programmer simply uses the \T{authors} measure
in \emph{both} the return type and the label of \T{getAuthors}.

\begin{figure}
\small
\begin{minipage}[t]{0.48\textwidth}
\begin{lifty}
showMyPapers ds client =
 let include p =  
  do auts <- getAuthors ds p
     return (elem client auts) in
 do 
  papers <- filterM include allPapers
  titles <- mapM (getTitle ds) papers
  print client (unlines titles)    
\end{lifty}
\vspace{.83cm}
\caption{\label{code:search-okay} A controller that displays all \T{client}'s papers (well-typed).}
\end{minipage}\hfill%
\begin{minipage}[t]{0.48\textwidth}
\begin{lifty}
showMyAccepts ds client =
 let include p =  
  do auts <- getAuthors ds p
     dec  <- getDecision ds p
     return (elem client auts 
             && dec == Accepted) in
 do 
  papers <- filterM include allPapers
  titles <- mapM (getTitle ds) papers
  print client (unlines titles)    
\end{lifty}
\caption{\label{code:search-leak} A controller that displays all \T{client}'s accepted papers (ill-typed).}
\end{minipage}

\end{figure}

\custompar{Checking self-referential policies via downgrading.}
Consider the controller \T{showMyPapers} in \autoref{code:search-okay},
which takes as argument a user \T{client}
and displays to them the titles of all the papers they authored.
To that end, \T{showMyPapers} filters the list of all paper identifiers \T{allPapers}
with a monadic predicate \T{include p},
which returns \vtrue iff \T{client} is an author of \T{p}.

At a first glance, this program should be rejected:
\T{showMyPapers} reads the author lists of every paper,
even those that \T{client} is not allowed to see.
Moreover, because of the self-referential policy,
the programmer finds themselves in a catch-22 situation:
in order to check whether the policy holds for a paper \T{p},
they must retrieve the author list of \T{p},
but that retrieval itself violates the policy!
Observe, however, that \T{showMyPapers} does not in fact leak anything to the user.
In particular, it does not allow \T{client} to distinguish between two data stores
that only differ in author lists that \T{client} is not allowed to see.
So, we would like \T{showMyPapers} to be well-typed, but that requires the type
checker to perform nontrivial reasoning about the values returned by \T{include
p}.

This is exactly where the \toolname's \T{downgrade} primitive comes in:
it allows the type checker to downgrade the input label on \T{include p},
because it always returns \vfalse for those papers that \T{client} is not allowed to see.
This is not a coincidence:
in fact, any (correctly implemented) runtime check of a self-referential policy
is well-typed (only) if it is wrapped in a \T{downgrade}.
In our example, the programmer does not use downgrading explicitly because it
is built into \toolname's \T{filterM} function,
giving it the following, very strong type:
\begin{lifty}[numbers=none]
filterM :: /\_a,i,f. (x: _a ->  TI {Bool|_v => f x} <f x && i>) 
                     -> [_a] -> TI [{a|f _v}] <i>
\end{lifty}
This type permits the filter predicate to have a higher label than the overall computation,
as long as for each list element \T{x}, 
the difference in labels \T{f x} is implied by the return value of the predicate.
This, in turns, allows the code in \autoref{code:search-okay} to type-check in
\toolname completely automatically.

\custompar{Information leaks through search.}
Now consider the controller \T{showMyAccepts} in \autoref{code:search-leak},
which is similar but only shows \T{client} their \emph{accepted} papers.
This code has an information leak of the similar nature as our original example in \autoref{code:edas-leak}:
the decision leaks to \T{client} through the list of paper titles before the reviewing in done.
This example is inspired by real-world leaks through search and recommendation functionality of data-centric applications
(see \autoref{sec:eval} for concrete examples). 

As expected, the \toolname type-checker identifies the input action \T{getDecision ds p} as the cause of the leak,
and replaces it with the same leak patch it generated for the EDAS leak.
With this patch, \T{showMyAccepts} always returns an empty list if called before reviewing is done;
the programmer deems this behavior acceptable and therefore accepts the patch.

\custompar{Attacker model.}
\toolname'a attacker model is standard, and similar to that of sequential
\LIO~\cite{StefanMMR17}.
In particular, we assume a language-level attacker who supplies \T{TIO} actions
(which we ensure to be well-typed as \tio{\T{()}}{\top}{\bot}).
%
%
We assume that policies are correct (\ie the type annotations on input-output
actions) and that the underlying \toolname infrastructure (in particular the
type checker, SMT solver, runtime, operating system, etc.) are secure.
Like most previous work on language-level IFC, we consider side channels and
transient execution attacks out of scope.

\section{The Core Calculus}\label{sec:formal}

We now formalize a core language \corelg, 
which captures the essence of \toolname's IFC mechanism.
We first present its syntax, as well as dynamic and static semantics.
The main goal of this section is to prove a noninterference guarantee,
which we accomplish by reduction to \LIO~\cite{StefanMMR17}.

\subsection{Syntax of \corelg}\label{sec:formal:syntax}

\begin{figure*}
\newcommand\subhead[1]{\multicolumn{3}{c}{\textbf{#1}}}
\newcommand\descr[1]{\text{\scriptsize{\emph{#1}}}}
\begin{minipage}[t]{.45\textwidth}
$$
\begin{array}{l@{~}ll}
\subhead{Refinements} \\
r, l ::=& \true  \mid  \false  \mid x \mid & \descr{Ref. terms}\\
     & \mid  \neg r \mid r \land r \mid a\\
\subhead{Types} \\
B ::=& \T{()} \mid \T{Bool} \mid \T{User}                     & \descr{Base types}\\        
T ::=& \{B \mid r\}  \mid  \funT{x}{T_1}{T_2}                 & \descr{Types}\\
  &  \mid \tio{T}{l_1}{l_2}\\
  &  \mid \fld{T}{l}
\end{array}
$$
\end{minipage}%
\begin{minipage}[t]{.55\textwidth}
$$
\begin{array}{l@{~}ll}
\subhead{Programs} \\
b &::= \unit \mid \vtrue \mid \vfalse \mid \vuser_1 \mid \vuser_2 \mid\ldots         & \descr{Base Values}\\
f &::= \vfield_1 \mid \vfield_2 \mid\ldots                                           & \descr{Fields}\\   
v &::= b \mid f \mid \lambda x.~t \mid \tioval{t}                                    & \descr{Values}\\
t &::= v \mid  x  \mid  t_1 ~ t_2 \mid \kwif ~ t_1 ~ \kwthen ~ t_2 ~ \kwelse ~ t_3   & \descr{Terms}\\
  &\quad \mid \kwreturn ~ t ~ \mid \kwbind~t_1~t_2\\ 
  &\quad \mid \kwdown~t_1~t_2 \\
  &\quad \mid \kwget ~ t \mid \kwset~t_1~t_2
\end{array}
$$
\end{minipage}
\caption{\label{formalism:syntax}
  Syntax of the core language \corelg.}
\end{figure*}

\corelg is a pure call-by-name $\lambda$-calculus,
equipped with refinement types and IFC constructs.
We summarize its syntax in \autoref{formalism:syntax}.

\custompar{Refinements.}
As is common in refinement types literature~\cite{RondonKaJh08,VazouSeJhVyJo14,PolikarpovaKuSo16},
\corelg distinguishes between program terms $t$
and refinement terms $r$, used inside types;
for readability, we use the meta-variable $l$ instead of $r$ 
for those refinement terms that represent labels.
We assume a syntactic category of atomic refinement terms $a$
drawn from a first-order theory.
For example, for the theory of equality and uninterpreted functions,
$a$ consists of equalities such as $x = \T{phase}~ds$.
Our formalization is agnostic to the choice of theory,
as long as validity of universal formulas of the form $\forall x_1,\ldots,x_n.r$ is decidable%
\footnote{Our implementation uses the theory of arrays, uninterpreted functions, and linear integer arithmetic.}.


\custompar{Types.}
The base types $B$ in \corelg include the unit type, the booleans, and the type of users.
%
As is standard in refinement types systems,
types include refined base types and function types.
In a refined base types $\{B\mid r\}$,
$r$ is a refinement predicate over the program variables and a special \emph{value variable} $\nu$, 
which denotes the bound variable of the type.
We sometimes write $B$ as a shortcut for $\{B\mid \true\}$.
%
Although the \toolname implementation supports dependent function types of the form $x\colon T_1 \to T_2$
(where the refinement of $T_2$ can mention the argument $x$),
and indeed we use them to specify policies in data-centric applications,
they are not central to our formalization, 
and hence the dependency is omitted from \corelg for simplicity.

One non-standard feature of \corelg is the type of sensitive computations \tio{T}{l_1}{l_2},
indexed by the return type $T$, and input and output \emph{labels} $l_1$ and $l_2$.
Both labels are refinement predicates over the program variables and a special \emph{user variable} $\upsilon$.
Since it is convenient to think of labels as elements of a lattice, 
we define the following lattice-theoretic syntax for logical operations on labels:
$$
l_1 \Fjoin l_2 \triangleq l_1 \wedge l_2 \quad\quad 
l_1 \Fmeet l_2 \triangleq l_1 \vee l_2 \quad\quad 
\top \triangleq \false \quad\quad 
\bot \triangleq \true
$$

Another non-standard feature is a dedicated type of \emph{fields} \fld{T}{l},
which is used to model \toolname's atomic input/output actions
and intuitively represents a resource at security level $l$ that stores values of type $T$
(where $T$ is restricted to refined base types).

\custompar{Program terms.}
Base values $b$ include unit, booleans, and a finite set of user literals $\vuser_i$. 
We also assume a finite set of field literals $f$,
each of which has a fixed type and label,
denoted by $\typeof(f)$ and $\labelof(f)$, respectively.
We denote the set of all base values as $\bvalues$
and the set of all fields as $\fields$.
In addition to base values and fields,
values $v$ include lambda abstractions 
and monadic actions \tioval{t}.
Like in prior work~\cite{StefanMMR17,VassenaR16,buiras2015hlio}, 
the \tiocons data constructor is not part of the surface syntax.

Terms $t$ include values, variables, function application, conditionals,
as well as monadic primitives \T{return}, \T{bind}, and \T{downgrade}
introduced in \autoref{sec:overview}.
Atomic input/output actions are represented in \corelg as two universal actions
\kwget and \kwset parameterized by the field to read from or write to.
The \T{downgrade} construct in \corelg also takes an additional field argument,
whose role is simply to specify the label to downgrade to:
intuitively, $\kwdown~f~t$ downgrades the result of $t$ to $\labelof(f)$.
In the \toolname implementation this label is implicit and inferred by the type checker;
the reason \corelg needs to reify this label in the term language
will become clear in \autoref{sec:formal:proof}.
%
%
We omit recursion from \corelg, since it does not present distinct challenges
for static IFC.
The \toolname implementation supports recursion,
and uses refinement types to prove that all recursive functions terminate. 


\subsection{Dynamic Semantics of \corelg}\label{sec:formal:dynamic}

\begin{figure*}
\addtolength{\jot}{2mm}
\textbf{Evaluation}\quad$\boxed{\config{\store}{t} \bigeval \config{\store'}{v}}$
\smallskip
\begin{gather*}
  \inference[\textsc{ret}]
    {}
    {\config{\store}{\kwreturn~t} \bigeval \config{\store}{\tioval{t}}}
  \quad
  \inference[\textsc{down}]
    {\config{\store}{t_2} \bigeval \config{\store'}{\tioval{b}}}
    {\config{\store}{\kwdown~t_1~t_2} \bigeval \config{\store'}{\tioval{b}}}  
  \\
  \inference[\textsc{bind}]
    {\config{\store}{t_1} \bigeval \config{\store'}{\tioval{t_1'}} &
      \config{\store'}{t_2 ~ t_1'} \bigeval \config{\store''}{v}}
    {\config{\store}{\kwbind ~ t_1 ~ t_2} \bigeval \config{\store''}{v}}    
  \\
  \inference[\textsc{get}]
    {\config{\store}{t} \bigeval \config{\store}{f}}
    {\config{\store}{\kwget ~ t} \bigeval \config{\store}{\tioval{\store[f]}}}
  \quad
  \inference[\textsc{set}]
    {\config{\store}{t_1} \bigeval \config{\store}{f} & \config{\store}{t_2} \bigeval \config{\store}{b}}
    {\config{\store}{\kwset ~ t_1 ~ t_2} \bigeval \config{\store[f := b]}{\tioval{\unit}}}  
\end{gather*}
\caption{\label{formalism:operational}
  Big-step operational semantics of IFC constructs in \corelg (see \autoref{appendix:language} for the full semantics).}
\end{figure*}

To model input and output actions, we define the run-time behavior of \corelg
in terms of a \emph{store} $\store\colon \fields\to\bvalues$
that maps fields to base values.
%
%
A \emph{program configuration} \config{\store}{t} consists of a store and a term.
\autoref{formalism:operational} defines a big-step evaluation relation \bigeval on configurations.
The behavior of monadic primitives is mostly straightforward;
in particular, since \corelg only tracks labels statically,
there is no label propagation or access checks at run time.
The evaluation order of pure and monadic terms is mostly standard for a call-by-name calculus~\cite{tackling},
in particular:
\kwbind is strict in its first argument, as expected;
\kwset is strict to ensure that only values are written to the store.
\kwdown fully evaluates its second argument---including under the \tiocons constructor---%
but ignores its field argument;
the purpose of the field argument will be clear from the noninterference proof
in \autoref{sec:formal:proof}.
The rules for pure terms are standard and therefore omitted,
and the rule for \kwdown is slightly simplified for exposition
(the full set of rules can be found in \autoref{appendix:language}).
Note that evaluating a pure term---\ie a term of a non-\T{TIO} type---%
does not change the store.

%

\subsection{Static Semantics of \corelg}\label{sec:formal:static}

\begin{figure*}
\addtolength{\jot}{2mm}
\renewcommand\colon{{\,:\,}}  
\textbf{Well-formedness}\quad$\boxed{\env \vdash T}$
\begin{gather*}
  \inference[\textsc{wf-base}]
    {\env, \nu\colon B \vdash r : \T{Bool}}
    {\env \vdash \{B \mid r\}}
  \quad
  \inference[\textsc{wf-TIO}]
    {\env\vdash T & \env,\upsilon \colon\T{User} \vdash l_i \wedge l_o: \T{Bool}}
    {\env \vdash \tio{T}{l_i}{l_o} }
\end{gather*}

\medskip    

\textbf{Subtyping and Flow}\quad$\boxed{\env \vdash T \Subt T'}$\quad$\boxed{\env \vdash l \Flows l'}$
\begin{gather*}
  \inference[\textsc{$\Subt$-TIO}]
    {\env \vdash T_1 \Subt T_2 & \env \vdash l_1 \Flows l_2 & \env \vdash l_2' \Flows l_1'}
    {\env \vdash \tio{T_1}{l_1}{l_1'} \Subt \tio{T_2}{l_2}{l_2'}}
  \quad
  \inference[\textsc{flow}]
    {\env, \upsilon\colon\T{User} \vDash l' \Implies l}
    {\env \vdash l \Flows l'}
\end{gather*}

\medskip
\textbf{Typing}\quad$\boxed{\env \vdash t::T}$
\begin{gather*}
  \inference[\textsc{t-ret}]
    {\env \vdash t :: T}
    {\env \vdash \kwreturn ~ t :: \tio{T}{\bot}{\top}}
  \\  
  \inference[\textsc{t-bind}]
    {\env \vdash t_1 :: \tio{T_1}{l_1}{l_1'} & \env \vdash t_2 :: \funT{x}{T_1}{\tio{T_2}{l_2}{l_2'}} & \env \vdash l_1 \Flows l_2'}
    {\env \vdash \kwbind ~ t_1 ~ t_2 :: \tio{T_2}{l_1 \Fjoin l_2}{l_1' \Fmeet l_2'}}
  \\
  \inference[\textsc{t-down}]
    {\env \vdash t_1 :: \fld{\_}{l} & \env \vdash t_2 :: \tio{\{\T{Bool} \mid \nu \Implies r\}}{l \Fjoin r}{l'}}
    {\env \vdash \kwdown ~ t_1 ~ t_2 :: \tio{\T{Bool}}{l}{l'}}
  \\
  \inference[\textsc{t-get}]
    {\env \vdash t :: \fld{T}{l}}
    {\env \vdash \kwget ~ t :: \tio{T}{l}{\top}}
  \quad
  \inference[\textsc{t-set}]
    {\env \vdash t_1 :: \fld{T}{l} & \env \vdash t_2 :: T}    
    {\env \vdash \kwset ~ t_1 ~ t_2 :: \tio{\T{()}}{\bot}{l}}
\end{gather*}    

\caption{\label{formalism:types}
  Static semantics of IFC constructs in \corelg (see \autoref{appendix:language} for the full semantics).}
\end{figure*}

\autoref{formalism:types} shows a subset of typing rules for \corelg
that are relevant to IFC.
Other rules are standard for languages with decidable refinement types
and deferred to \autoref{appendix:language}.
In the figure, a \emph{typing environment} $\env ::= \epsilon \mid \env, x:T \mid \env, r$
maps variables to types 
and records \emph{path conditions} $r$,
which arise when checking conditional terms.

\custompar{Well-formedness.}
We show well-formedness rules for base and \T{TIO} types.
They rely on the auxiliary \emph{sorting} judgment $\env\vdash r: B$,
which depends on the underlying refinement logic, but necessarily checks that $r$ only mentions variables in $\env$.
The two rules formalize the distinction between logical refinements and labels:
the former can mention the value variable $\nu$ and the latter can mention the user variable $\upsilon$.
Well-formedness premises are implicit in all typing rules described below.

\custompar{Subtyping.}
%
The rule \textsc{$\Subt$-TIO} specifies that tagged types 
are covariant in their input label and contra-variant in the output label \wrt the \emph{can-flow-to} order on labels.
This order, $\env\vdash l \Flows l'$ is defined as the logical validity 
of reverse implication between label predicates,
under the assumptions stored in the environment
(which include path conditions and refinements on program variables).
For example, the judgment $a\colon \T{User}, b\colon \{\T{User} \mid \nu = a\} \vdash \upsilon = a \Flows \upsilon = b$ 
reduces to a formula $\forall a,b,\upsilon . b = a \wedge \upsilon = b \Implies \upsilon = a$,
and hence is valid.
Recall that under our assumptions on the refinement logic,
the validity of such formulas is decidable.
Note that for a given $\env$, the set of labels with $\Flows, \Fjoin, \Fmeet, \top, \bot$ forms a lattice.
%

Subtyping for base types and function types is standard.
Importantly, field types are \emph{invariant} in both the value type and the label
(since fields are used for reading and writing).

\custompar{Term typing.}
The rest of \autoref{formalism:types} defines
the typing judgment for program terms $\env \vdash t::T$.
The first three rules encode the same label propagation logic
we have introduced in \autoref{code:primitives};
the difference is that in \corelg the monadic primitives are built-in
and have custom typing rules,
whereas in the actual \toolname implementation they are represented as library functions with polymorphic types.
Note that we leverage the flexibility of custom typing rules 
to simplify the static semantic of \kwbind slightly:
here we add an explicit premise $\env \vdash l_1 \Flows l_2'$,
whereas in \autoref{code:primitives} we encode this implicitly by representing $l_2'$ as $l_1 \Fjoin p$.
The last two rules type the universal actions \kwget and \kwset
using the type of the corresponding field.

\subsection{Noninterference in \corelg}\label{sec:formal:proof}

In this section we show that \toolname programs cannot leak sensitive data by
proving \emph{noninterference} for the core calculus. 
%
Instead of proving noninterference from first principles,
we accomplish this by reducing \corelg to core \LIO, 
a language with dynamic IFC, whose noninterference proof has been mechanized in Coq~\cite{StefanMMR17}.
Our proof strategy is a as follows:
first, we present \emph{instrumented} operational semantics that adds dynamic IFC to \corelg;
next, we argue that the instrumented \corelg is a subset of core \LIO and hence exhibits noninterference;
finally, we show that well-typed terms behave equivalently under the original and instrumented semantics.

\begin{figure*}
\textbf{Instrumented Evaluation}\quad$\boxed{\config{\store,\instr{\ell_c}}{t} \liobigeval \config{\store',\instr{\ell'_c}}{v}}$
\smallskip
\begin{gather*}
  \inference[\textsc{i-ret}]
    {}
    {\config{\store,\instr{\ell_c}}{\kwreturn~t} \liobigeval \config{\store,\instr{\ell_c}}{\tioval{t}}}
  \\
  \inference[\textsc{i-bind}]
    {\config{\store,\instr{\ell_c}}{t_1} \liobigeval \config{\store',\instr{\ell'_c}}{\tioval{t_1'}} &
    \config{\store',\instr{\ell'_c}}{t_2 ~ t_1'} \liobigeval \config{\store'',\instr{\ell''_c}}{v}}
    {\config{\store,\instr{\ell_c}}{\kwbind ~ t_1 ~ t_2} \liobigeval \config{\store'',\instr{\ell''_c}}{v}}    
  \\
  \inference[\textsc{i-down-1}]
    {\config{\store,\instr{\ell_c}}{t_1} \liobigeval \config{\store,\instr{\ell_c}}{f} &
     \config{\store,\instr{\ell_c}}{t_2} \liobigeval \config{\store',\instr{\ell'_c}}{\tioval{b}} & 
     \instr{\sem{\labelof(f)} = \ell} &
     \instr{\ell'_c \Flows \ell \Fjoin \ell_c}}
    {\config{\store,\instr{\ell_c}}{\kwdown~t_1~t_2} \liobigeval \config{\store',\instr{\ell\Fjoin\ell_c}}{\tioval{b}}}
  \\
  \inference[\textsc{i-down-2}]    
    {\config{\store,\instr{\ell_c}}{t_1} \liobigeval \config{\store,\instr{\ell_c}}{f} &
     \config{\store,\instr{\ell_c}}{t_2} \liobigeval \config{\store',\instr{\ell'_c}}{\tioval{b}} & 
     \instr{\sem{\labelof(f)} = \ell} &
     \instr{\ell'_c \not\Flows \ell \Fjoin \ell_c}}
    {\config{\store,\instr{\ell_c}}{\kwdown~t_1~t_2} \liobigeval \config{\store',\instr{\ell\Fjoin\ell_c}}{\tioval{\vfalse}}}
  \\
  \inference[\textsc{i-get}]
    {\config{\store,\instr{\ell_c}}{t} \bigeval \config{\store,\instr{\ell_c}}{f} & \instr{\sem{\labelof(f)} = \ell}}
    {\config{\store,\instr{\ell_c}}{\kwget ~ t} \liobigeval \config{\store,\instr{\ell\Fjoin \ell_c}}{\tioval{\store[f]}}}
  \\
  \inference[\textsc{i-set}]
    {\config{\store,\instr{\ell_c}}{t_1} \bigeval \config{\store,\instr{\ell_c}}{f} &
     \config{\store,\instr{\ell_c}}{t_2} \liobigeval \config{\store,\instr{\ell_c}}{b} & 
     \instr{\ell_c \Flows \sem{\labelof(f)}}}
    {\config{\store,\instr{\ell_c}}{\kwset ~ t_1 ~ t_2} \liobigeval \config{\store[f := b],\instr{\ell_c}}{\tioval{\unit}}}
\end{gather*}
\caption{\label{formalism:instrumented}
  Instrumented operational semantics of IFC constructs in \corelg (see \autoref{appendix:language} for the full semantics).}
\end{figure*}
 
\subsubsection{Instrumented Semantics}



An \emph{instrumented configuration} $k$ is a triple \config{\store,\ell_c}{t}, where $\ell_c$ is the \emph{current (program counter) label}.
Intuitively, $\ell_c$ starts out at $\sem{\bot}$ and then gradually rises as the evaluation progresses,
keeping track of the most sensitive field that the computation has read so far
and blocking output to any field not above $\ell_c$.
In the interest of clarity, we introduce a distinct syntactic category $\ell$ of \emph{run-time labels},
which are labels that do not mention any variables except $\upsilon$;
the \emph{can-flow-to} judgment between run-time labels does not require an environment,
so we write it simply as $\ell \Flows \ell'$.
Note that all labels of field literals $\labelof(f)$ are naturally run-time labels.

\autoref{formalism:instrumented} defines a big-step evaluation relation \liobigeval on instrumented configurations.
The rules for pure terms keep the current label intact and are omitted
(the full set of rules can be found in \autoref{appendix:language}).
The core mechanism for propagating and checking run-time labels 
is captured in the rules \textsc{i-get}, \textsc{i-set}, and \textsc{i-bind}.
The rule \textsc{i-get} raises the current label by $\ell$---the label of the field being read.
The premise of \textsc{i-set} checks that the current label can flow to the label of the field being written.
Finally the rule \textsc{i-bind} uses the final label of the first action as the starting label of the second action.

The most interesting part of the instrumented semantics is the behavior of $\kwdown~t_1~t_2$,
expressed in \textsc{i-down-1} and \textsc{i-down-2}.
Both of these rules start by fully evaluating $t_1$ to $f$ 
(which changes neither the store nor the current label, since $t_1$ is a pure term)
and then $t_2$ to \tioval{b} 
(\ie either \tioval{\vtrue} or \tioval{\vfalse}).
The latter evaluation raises the current label to $\ell_c'$.
Instead of adopting $\ell_c'$ as the new current label, however,
both rules \emph{downgrade} it to $\ell \Fjoin \ell_c$,
effectively only raising the current label by the manifest label $\sem{\labelof(f)}$ of the \kwdown operation.
But won't such downgrading leak information at level $\ell_c'$ through $b$?
This is where the \textsc{i-down-2} comes in:
if in fact $\ell_c'$ can not flow to the downgraded label,
then the true value of $b$ is discarded and \vfalse is returned instead.

\subsubsection{From Instrumented Semantics to \LIO}

We argue that the semantics in \autoref{formalism:instrumented} is equivalent to a subset of sequential \LIO with references, 
as defined in~\cite{StefanMMR17}.
For pure terms, as well as \kwreturn and \kwbind the equivalence is straightforward by comparing the evaluation rules.
The remaining primitives \kwget, \kwset, and \kwdown can be encoded in \LIO as follows:
\begin{lifty}[numbers=none]
get f         `$\equiv$` readLIORef f
set f t       `$\equiv$` writeLIORef f t
downgrade f t `$\equiv$` do lc <- getLabel                    
                    lb <- toLabeled ((labelOf f) `$\Fjoin$` lc) t
                    catchLIO (unlabel lb) (\_ -> return False)
\end{lifty}
Both \kwget and \kwset simply read and write a reference \T{f},
labeled with $\T{l} = \labelof(\T{f})$
(created at the start of the program with \T{newLIORef l}).
The more interesting case is \kwdown.
The \T{toLabeled} primitive returns a labeled value---whose label is \T{l}~$\Fjoin$~\T{lc}---that
either contains the result of \T{t} or a ``delayed'' exception if
\T{t} reads data more sensitive than \T{l}~$\Fjoin$~\T{lc}.
\T{unlabel} raises the current label to \T{l}~$\Fjoin$~\T{lc} and either
returns the value computed within the \T{toLabeled} block or throws the delayed
exception---hence the need to catch the exception and return \vfalse.

Since noninterference in \LIO has been proven \wrt an arbitrary security lattice,
the instrumented \corelg inherits its guarantee.
Noninterference is formalized in terms of \emph{$\ell$-equivalence} on instrumented configurations,
which is formally defined in \autoref{appendix:language}.
Intuitively, two stores are considered $\ell$-equivalent if they only differ in fields whose label is not below $\ell$.

\begin{lemma}[Noninterference of instrumented semantics] \label{formal:lio}
Instrumented evaluation from $\ell$-equivalent configurations leads to $\ell$-equivalent configurations:
if $k_1 \lequiv{\ell} k_2$,
$k_1 \liobigeval k_1'$, 
and $k_2 \liobigeval k_2'$,
then $k_1' \lequiv{\ell} k_2'$
\end{lemma}

\subsubsection{Simulation}

The core of our noninterference argument is a proof that
instrumented execution simulates original execution for well-typed terms.
The full proofs can be found in \autoref{appendix:language};
here we only state the key lemma and give the intuition for the proof.

\begin{lemma}[Simulation] \label{formalism:simulation}
If $\epsilon\vdash t :: \tio{T}{\ell_i}{\ell_o}$ and $\config{\store}{t} \bigeval \config{\store'}{v}$,
then for any $\ell_c \Flows \ell_o$, there exists a new current label $\ell_c'$ such that
\begin{inparaenum}[(1)] 
\item $\config{\store,\instr{\ell_c}}{t} \liobigeval \config{\store',\instr{\ell'_c}}{v}$,
\item $\ell_c' \Flows \ell_c \Fjoin \ell_i$,
\item $\epsilon\vdash v :: \tio{T}{\ell_i}{\ell_o}$.
\end{inparaenum}
\end{lemma}

Intuitively, this lemma says that executing a well-typed monadic term $t$
from a configuration where the current label $\ell_c$ is not too-high (with respect to the static output label)
leads to the same result under the instrumented semantics (1).
In addition, we also show that instrumented execution would only raise the current label
by the static input label of the computation (2).
The proof is by induction on the derivation of big-step evaluation.
Interesting cases include \kwset, where we show that the runtime check never fails (this follows from $\ell_c \Flows \sem{\ell_o}$);
\kwbind, where we show that after executing the first action, the current label 
remains below the output label of the second action (this follows from property (2) and the last premise of \textsc{t-bind});
and \kwdown, where we show that whenever \textsc{i-down-2} applies, $b$ is $\vfalse$ anyway.

Finally we can combine Lemma~\ref{formal:lio},
Lemma~\ref{formalism:simulation}, and \LIO's noninterference proof to show
noninterference of \corelg programs:

\begin{theorem}[Noninterference for \corelg] \label{formal:noninterference}
Evaluating a computation $t$ statically visible to $\ell$
from $\ell$-equivalent stores leads to $\ell$-equivalent stores:
If $\store_1 \lequiv{\ell} \store_2$,
$\epsilon\vdash t :: \tio{T}{\ell}{\_}$,
$\config{\store_1}{t} \bigeval \config{\store_1'}{v_1}$,
and $\config{\store_2}{t} \bigeval \config{\store_2'}{v_2}$,
we have $v_1 = v_2$ and $\store_1' \lequiv{\ell} \store_2'$.
\end{theorem}

Technically, our noninterference guarantee is \emph{termination-insensitive}:
we state that final configurations are equivalent
only as long as both initial configurations evaluate to a value.
However in \corelg this is not an issue, 
since \emph{all} well-typed programs evaluate to a value:
progress and termination can be shown by a straightforward extension
of proofs of these properties for simply-typed lambda calculus.
More interestingly, the full \toolname language also enjoys this property:
although it supports recursion,
it uses refinement types to prove that all recursive calls terminate.

\section{Leak Repair in \corelg}\label{sec:synthesis}

\begin{figure}
\small
\begin{minipage}{.52\textwidth}
  \begin{algorithmic}[1]
    \Function{Enforce}{$\env, t$}
      \Let{$\hat{t}$}{\Call{Localize}{$\env, t$}}\label{alg:localize}
      \For{$\cast{T_a}{T_e}t_a \in \hat{t}$}
        \Let{$t_p$}{\Call{Generate}{$x\colon T_a, \env, T_e$}$[x \mapsto t_a]$}\label{alg:gen}
        \Let{$\hat{t}$}{$\hat{t}[\cast{T_a}{T_e}t_a ~ \mapsto ~ t_p]$}
      \EndFor       
    \EndFunction
    \Function{Localize}{$\env, t$}
      \Let{$\hat{t}$}{$t$}
      \Let{$(\mathit{rules},\mathit{queries})$}{\Call{CHC}{$\env \vdash t :: \tio{\T{()}}{\top}{\bot}$}}\label{alg:gen-chc}
      \Let{$\mathcal{A}$}{\Call{HornSolver}{$\mathit{rules}$}}\label{alg:unfold}
      \For{$Q \gets \mathit{queries} \mid \mathcal{A}\not\vDash Q$}
        \Let{$(t_a, \mathit{sub})$}{$\mathit{source}(Q)$}
        \If{$\mathit{sub} = \ti{B}{i} \Subt \ti{B}{l}$}
          \Let{$\hat{t}$}{$[t_a \mapsto \cast{\ti{B}{i}}{\ti{B}{\mathcal{A}[l]}}t_a]\hat{t}$}
        \Else\ \textbf{fail}  
        \EndIf
      \EndFor      
    \EndFunction
  \end{algorithmic}
\end{minipage}%
\begin{minipage}{.48\textwidth}
  \begin{algorithmic}[1]
    \setcounter{ALG@line}{14}
    \Function{Generate}{$\env_R, \env_G, \ti{B}{l}$}\label{alg:gen-start}
      \Let{$\env_R$}{$\env_R \cup$ $\mathcal{R}$}
      \Let{$\mathit{branches}$}{\Call{SynthAll}{$\env_R\vdash \T{??}::\ti{B}{\top}$}}\label{alg:synt1}
      \Let{$\mathit{conds}$}{\Call{Abduce}{$\env_G,\T{??}\vdash t_b :: \ti{B}{l}$} \\\quad\quad\quad\quad\quad\quad \textbf{for} $t_b \gets \mathit{branches}$}\label{alg:abduce}
      \Let{$((t_d, r_g) : \mathit{guarded})$}{\Call{Sort}{$\mathit{branches}, \mathit{conds}$}}\label{alg:sort}
      \If{$r_g \liff \true$}\label{alg:check}
        \Let{$t$}{$t_d$}\label{alg:def}
      \Else\ \textbf{fail}\label{alg:fail}      
      \EndIf            
      \For{$(r_g, t_b) \gets \mathit{guarded}$}
        \Let{$T_g$}{$\ti{\{\T{Bool}\mid\nu \Leftrightarrow r_g\}}{l}$}
        \Let{$t_g$}{\Call{Synth}{$\env_G\vdash \T{??}::T_g$}}\label{alg:synt2}
        \Let{$t$}{$\kwbind~t_g~(\lambda y.\T{if}~y~\T{then}~t_b~\T{else}~t)$}\label{alg:combine}        
      \EndFor 
      \State \Return{$t$}
    \EndFunction    
  \end{algorithmic}
\end{minipage}  
\caption{Leak repair algorithm.}\label{alg:enforce}
\end{figure}

We now formalize \toolname's leak repair mechanism for the core calculus \corelg.
\autoref{alg:enforce} shows the pseudocode of the algorithm \textproc{Enforce},
which performs type-checking and repair of an individual controller function.
More precisely, the algorithm takes as input a typing environment $\env$ and a program $t$,
and determines whether $t$ can be patched to produce a well-typed \T{TIO} computation,
\ie a term $t'$ such that $\env \vdash t' :: \tio{\T{()}}{\top}{\bot}$.
The algorithm proceeds in two steps.
First, procedure \textproc{Localize} identifies unsafe terms (line~\ref{alg:localize}),
replacing them with \emph{type casts} to produce a ``program with holes'' $\hat{t}$
(\autoref{sec:synthesis:localization}).
Then, the algorithm replaces each type cast in $\hat{t}$ with an appropriate patch, 
generated by the procedure \textproc{Generate}
(\autoref{sec:synthesis:generation}).

\subsection{Leak Localization}\label{sec:synthesis:localization}

\begin{figure*}
\addtolength{\jot}{2mm}
\begin{gather*}
  \inference[\textsc{t-cast}]
    { }
    {\env \vdash \cast{T}{T'} :: T\to T'}
  \quad  
  \inference[\textsc{l-get}]
    {\env \vdash \kwget ~ t :: T & T = \tio{B}{l}{\top} & T' = \tio{B}{l'}{\top}}
    {\env \vdash \kwget ~ t \casts \cast{T}{T'}~(\kwget ~ t) :: T'}
\end{gather*}    

\caption{\label{synth:casts}
  Cast typing and cast insertion for \corelg.}
\end{figure*}

\custompar{Type casts.}
For the purpose of leak localization, we extend the values of \corelg with type casts:
\[v ::= \cdots  \mid  \cast{T}{T'}\]
Statically, our casts are similar to those in prior work~\cite{KnowlesF10};
in particular, the cast $\cast{T}{T'}$ has type $T\to T'$,
as indicated in \autoref{synth:casts}.
However, the dynamic semantics of casts in \corelg is undefined:
casts are inserted solely for the purpose of leak localization,
and, if repair succeeds, are completely eliminated.
We restrict the notion of \emph{type-safe} \corelg programs to those
that are well-typed are \emph{free of type casts}.

\custompar{Cast insertion.}
Declaratively, leak localization can be formalized using a cast insertion judgment $\env\vdash t \casts \hat{t} :: T$,
which informally means that inserting type casts into term $t$ can yield a term $\hat{t}$ of type $T$.
Unlike prior work, our cast insertion is specific to IFC and our intended repair strategy%
---guarding and redacting unsafe input actions.
As a result, we only allow inserting casts around \kwget expressions, 
as shown in the \textsc{l-get} rule in~\autoref{synth:casts};
for all other terms the judgment is defined homomorphically.
Furthermore, the rule \textsc{l-get} imposes two important restrictions:
\begin{inparaenum}[(1)]
\item the cast can only change the input label of the action (intuitively, it downgrades $l$ into $l'$), and
\item the cast must be \emph{functionally oblivious}: the result type $\{B\mid r\}$ of \T{TIO} must have a trivial refinement, \ie $r = \true$.
\end{inparaenum}
As we explain below, these restrictions are crucial for the efficiency of repair,
and although they introduce incompleteness, we found them to work well in practice.
In particular, functionally oblivious casts make sense in our use case
because patch generation will redact the input action anyway,
so we are unlikely to be able to satisfy any functional property $r$ of the original action.

\custompar{Minimal sound localizations.}
Using the typing and cast insertion rules, 
we can show that if $\env\vdash t \casts \hat{t} :: T$ then $\env\vdash\hat{t} :: T$.
We refer to $\hat{t}$ as a \emph{sound localization} of $t$ at type $T$ in $\env$.
A sound localization $\hat{t}$ of $t$ is also \emph{minimal},
if replacing any $T_i'$ in $\cast{T_i}{T_i'}$ in $\hat{t}$ with its supertype
prevents $\hat{t}$ from type-checking against $T$.
The following lemma follows directly from well-typing of $\hat{t}$:

\begin{lemma}[Localization]\label{synthesis:sound-loc}
  If $\env\vdash t \casts \hat{t} :: T$,
  replacing each subterm of the form $\cast{T_i}{T_i'}~t_i$ in $\hat{t}$
  with a type-safe term of type $T_i'$ yields a type-safe program.
\end{lemma}

Once a sound localization is found,
this lemma enables patch generation to proceed \emph{independently} for each type cast
(taking the type $T_i'$ as the expected type).
If the localization is also minimal, patch generation has the highest chance to succeed.
Hence the goal of the leak localization algorithm 
is to find a minimal sound localization of $t$ at type $\tio{\T{()}}{\top}{\bot}$ in $\Gamma$.

In a general-purpose type-driven repair setting,
a term might have many minimal sound localizations,
and the repair engine would have to explore all of them,
until it finds one where all type casts can be patched,
leading to inefficiency.
For our domain-specific repair strategy, however,
there is no need to search through localizations:
in fact, given the restrictions we introduced on cast insertion,
any \corelg term has \emph{at most one} minimal sound localization 
(up to equivalence of refinement terms). 

To see why, recall that sound localizations can only differ in
expected labels of atomic input actions.
From the typing rules we know that in the typing derivation of $\hat{t}$,
the expected label $l'$ can only appear on the right-hand side of an implication $\Gamma\vDash l \Implies l'$;
hence the weakest expected type for each cast can be chosen independently:
it is the type with the highest label $l'$ that satisfies the above constraint.
If we omitted the requirement that casts be functionally oblivious
and allowed the expected type to be any \tio{\{B\mid r\}}{l'}{\bot},
the uniqueness property would be violated.
This is because, unlike labels, refinements $r$ can appear in a typing derivation as environment assumptions.
Hence cast insertion would have a trade-off:
picking a stronger type for one action (with a stronger $r$)
might validate a choice of a weaker type for another action. 

\custompar{Inferring the localization.}
Given the restrictions outlined above, 
finding the minimal sound localization for a \corelg term,
amounts to inferring the highest expected label for each unsafe access.
Procedure \textproc{Localize} formalizes our new algorithm
that infers expected labels efficiently during type checking.
It first uses the \corelg typing and subtyping rules
to reduce the problem of checking the source program $t$
to a system of \emph{constrained Horn clauses} (CHCs) over unknown refinements and labels.
As we explained in \autoref{sec:overview},
CHCs can be divided into \emph{rules} and \emph{queries}.
In line~\ref{alg:unfold}, we use an existing CHC solver~\cite{CosmanJ17}
to obtain the \emph{strongest assignment} $\mathcal{A}$
of refinement terms to unknowns.
If this assignment satisfies all the queries,
then $t$ is well-typed, and \textproc{Localize} terminates without modifying it.
Otherwise, for each query $Q$ that does not hold under $\mathcal{A}$,
we obtain its \emph{source},
\ie the term $t_a$ and the subtyping constraint that generated the query.
Now if the query was generating by a can-flow check from the input label $i$ of term $t_a$
to some (possibly unknown) label $l$,
then we insert a type-cast around $t_a$,
taking the expected label to be $\mathcal{A}[l]$,
\ie the valuation of $l$ in $\mathcal{A}$.
If, on the other hand, $Q$ is not derived from a can-flow check,
then the type error is not caused by an information leak,
hence \textproc{Localize} fails.



\subsection{Patch Generation}\label{sec:synthesis:generation}

Next, we describe how our algorithm replaces a type-cast $\cast{T_a}{T_e}~t_a$ with a patch term $t_p$ of the expected type $T_e$,
using the patch generation procedure \textproc{Generate} (line~\ref{alg:gen}).
\textproc{Generate} implements a domain-specific synthesis strategy:
first, it generates a list of \emph{branches},
which return the original term redacted to a different extent;
then, for each branch, it infers an optimal \emph{guard} (a policy check)
that makes the branch satisfy the expected type;
finally, it constructs the patch by arranging the properly guarded branches into a (monadic) conditional.

\custompar{Synthesis of branches.}
Given \ti{B}{l} as the goal type,
\textproc{Generate} first uses \synquid~\cite{PolikarpovaKuSo16} to synthesize the set of all terms up to certain size
of type \ti{B}{\top}, \ie with the right content type, but with no restriction on the label
(line~\ref{alg:synt1}).
Branches are generated in a restricted environment $\env_R$,
which contains only the original faulty term and a small set of \emph{redaction functions} $\mathcal{R}$.
This set is specified by the programmer,
and typically includes a ``default value'' of each type,
but may also include \eg functions that sanitize strings
(we show examples in \autoref{sec:eval}).
This restriction gives the user control over the space of patches
and also makes the synthesis more efficient.

\custompar{Synthesis of guards.}
Consider a branch $t_b$ synthesized by \textproc{SynthAll}:
its actual, strongest type is some \ti{B}{l'},
while the expected type of the patch is \ti{B}{l}.
\textproc{Generate} now attempts to synthesize the optimal guard that would make $t_b$ respect the expected type.
At a high level, this guard must be logically equivalent to the weakest refinement formula $r_g$, such that
\begin{inparaenum}[(1)]
\item $\env_G, r_g \vdash l' \Flows l$ and
\item $\env_G \vdash r_g : \T{Bool}$ (\ie $r_g$ does not mention the user variable $\upsilon$).
\end{inparaenum}
This formula can be inferred using existing techniques,
such as logical abduction~\cite{DilligD13}.
In particular, \textproc{Generate} relies on \synquid's \emph{liquid abduction} mechanism
to infer $r_g$ for each branch in line~\ref{alg:abduce}.

In line~\ref{alg:sort} we topologically sort the branches according to their abduced conditions, from weakest to strongest
(\ie in the reverse order of how they are going to appear in the program).
In line~\ref{alg:check}, 
we check that the first branch can be used as the \emph{default branch}, \ie it is correct unconditionally.
This property is always satisfied as long as $\env_R$ contains a pure value $b$ of type $B$,
in which case $\kwreturn~b$ is a valid default branch.

The main challenge of guard synthesis, 
is that the guard itself must be monadic, since it might need to retrieve and compute over some data from the store.
Since the data it retrieves might itself be sensitive,
we need to ensure that two conditions are satisfied
\begin{inparaenum}[(1)]
\item \emph{functional correctness:} the guard returns a value equivalent to $r_g$, and
\item \emph{no leaky enforcement}: the input label of the guard itself may flow to the expected label $l$ of the patch.
\end{inparaenum}
To ensure both conditions, we again use \synquid,
this time with the goal type \ti{\{\T{Bool}\mid\nu \liff r_g\}}{l}
to synthesize the guard.


\begin{lemma}[Safe patch generation]\label{synthesis:sound-generate}
  If \textproc{Generate} succeeds, it produces a type-safe term of the expected type \ti{B}{l}.
\end{lemma}
Assuming correctness of \textproc{Synth} and \textproc{Abduce},
we can use the typing rules of \autoref{sec:formal} 
to show that the invariant $\env_R \cup \env_G \vdash\mathit{patch}::\ti{B}{l}$
is established in line~\ref{alg:def} and maintained in line~\ref{alg:combine}.
In particular, the type of the bound variable $y$ in line~\ref{alg:combine} is $\{\nu:\T{Bool}\mid\nu \liff r_g\}$,
hence, \T{then} branch is checked under the path condition $r_g \liff \true$.
Since $r_g$ is the result of abduction, we know that $\env_G, r_g \vdash \ti{B}{l'} \Subt \ti{B}{l}$,
and hence $\env_G, r_g \vdash b :: \ti{B}{l}$.


\subsection{Guarantees and Limitations}\label{sec:synthesis:guarantees}

In this section we summarize the soundness guarantee of leak repair in \corelg
and then discuss the limitations on its completeness and minimality.

\begin{theorem}[Soundness of leak repair]\label{synthesis:soundness}
  If procedure \textproc{Enforce} succeeds, it produces a program that satisfies noninterference.
\end{theorem}
This is straightforward by combining Lemmas~\ref{synthesis:sound-loc} and \ref{synthesis:sound-generate}
with Theorem~\ref{formal:noninterference}.

\custompar{Completeness.}
When does procedure \textproc{Enforce} fail?
\textproc{Localize} fails when it cannot find a safe localization satisfying our domain-specific restrictions,
which happens if 
\begin{inparaenum}[(1)]
\item the program contains an error unrelated to information flow, or
\item the program depends on a functional property of an unsafe input action we want to redact.
\end{inparaenum}
We consider both of these cases out of scope of our domain-specific repair algorithm.
\textproc{Generate} can fail in lines~\ref{alg:fail} and \ref{alg:synt2}. 
The first failure indicates that $\env_R$ does not contain any sufficiently public terms;
in this case, \toolname prompts the programmer to add a default value of an appropriate type.
The second failure happens when no guard satisfies both functional and security requirements;
this commonly indicates that the policy is \emph{not enforceable} without leaking some other sensitive information.
For instance, in the EDAS leak example,
if the programmer declared ``phase'' to be only visible to the program chair,
no program could precisely check the policy on ``decision'' without leaking the information about ``phase''.
In this case, \toolname prompts the programmer to change the policies
in a way that respects dependencies between sensitive fields
(\ie to make the policy on ``decision'' at least as restrictive as the one on ``phase'').

\custompar{Minimality.}
Ideally, we would like to show that the changes made by \textproc{Enforce} are minimal:
in any execution where $t$ did not cause a leak, $t'$ would output the same values as $t$.
Unfortunately, this is not true:
\textproc{Enforce} is \emph{conservative} and might hide more information than is strictly necessary.
The reason for the imprecision is two-fold:
\begin{inparaenum}[(1)]
\item the minimal localization inferred by \textproc{Localize} might over-approximate the actual runtime label of output actions
due to imprecisions of refinement type inference; and  
\item the guard condition $r_g$ abduced by \textproc{Generate} might be overly strong
due to the limitations of the abduction engine.
\end{inparaenum}
In the latter case, the programmer can provide a more precise guard manually;
the former is a fundamental limitation of all static IFC systems.


\section{Evaluation}\label{sec:eval}

\begin{table}
\renewcommand\d[2]{\lookupPut{#1}{#2}}
\d{micro01}{\T{EDAS}}
\d{micro02}{\T{EDAS-Multiple}}
\d{micro03}{\T{EDAS-Self-Ref}}
\d{micro04}{\T{Search}}
\d{micro05}{\T{Sort}}
\d{micro06}{\T{Broadcast}}
\d{micro07}{\T{HotCRP}}
\d{micro08}{\T{AirBnB}}
\d{micro09}{\T{Instagram}}
\renewcommand\d[1]{\lookupGet{#1}} 
\centering
\renewcommand\arraystretch{1}
\newcommand\st{\rule{0pt}{1em}} 
\newcommand\vl[1]{\multicolumn{1}{c|}{#1}}
%
\scalebox{0.80}{
\begin{tabular}{rl|rr|rrr}
\toprule
 \multicolumn{2}{l|}{\multirow{2}{*}{\textbf{Benchmark}}} & \multicolumn{2}{c|}{\textbf{Code size (AST nodes)}} & \multicolumn{3}{c}{\textbf{Time}} \\
  
  & &  Original  &   \toolname     &  Localize\!\!            &     Generate\!\!  &  ~~Total  \\
\midrule
1   & \d{micro01}             &        66  &      25  &    0.1s  &      0.1s  &      0.3s  \\
2   & \d{micro02}             &        87  &      50  &    0.3s  &      0.3s  &      0.6s  \\
3   & \d{micro03}             &        87  &      76  &    0.3s  &      0.9s  &      1.2s  \\
4   & \d{micro04}             &        76  &      25  &    0.8s  &      0.2s  &      1.0s  \\
5   & \d{micro05}             &        64  &      58  &    0.5s  &      0.6s  &      1.1s  \\
6   & \d{micro06}             &        27  &      25  &    0.1s  &      0.2s  &      0.2s  \\
7   & \d{micro07}             &        68  &      22  &    0.5s  &      0.0s  &      0.5s  \\
8   & \d{micro08}             &        52  &      33  &    0.2s  &      0.1s  &      0.4s  \\
9   & \d{micro09}             &        73  &      42  &    0.5s  &      0.9s  &      1.4s  \\
\bottomrule
\end{tabular}
}
\caption{\label{eval:benchmarks:micro}
  Microbenchmarks.}
  \vspace{-.5cm}
\end{table}

\begin{table*}
\renewcommand\d[2]{\lookupPut{#1}{#2}}
\newcommand\vl[1]{\multicolumn{1}{c|}{#1}}
\newcommand\st{\rule{0pt}{1em}} 
\renewcommand\arraystretch{1}
\bigskip
  \noindent \hspace{1cm} {\sffamily\footnotesize (a) Conference Management System} \hfill 
  \hbox{\scalebox{0.80}{\fbox{Policy size (AST nodes):  247}} \hspace{1cm}} 
\d{registerUser}{Register user}
\d{usersView}{View users}
\d{submitForm}{Paper submission}
\d{searchForm}{Search papers}
\d{paperView}{Show paper record}
\d{reviewsView}{Show reviews for paper}
\d{profileViewGet}{User profile: GET}
\d{profileViewPost}{User profile: POST}
\d{submitReviewViewPost}{Submit review}
\d{assignReviewersView}{Assign reviewers}
\renewcommand\d[1]{\lookupGet{#1}} 
\scalebox{0.80}{
\begin{tabular}{l|rrr|rrrr}
\toprule
  \multirow{2}{*}{\textbf{Benchmark}}  & \multicolumn{3}{c|}{\textbf{Code size (AST nodes)}} & \multicolumn{4}{c}{\textbf{Time}} \\
&  Original  &  Manual  & \toolname &   Verify\st      &  Localize            &     Generate  &  Total  \\
\midrule
\d{registerUser}                &        22  &       0  &       0  &      0.1s  &      0.1s  &      0.0s  &      0.1s  \\
\d{usersView}                   &        26  &      16  &      25  &      0.5s  &      0.2s  &      0.2s  &      0.4s  \\
\d{submitForm}                  &        65  &       0  &       0  &      2.0s  &      2.2s  &      0.0s  &      2.2s  \\
\d{searchForm}                  &       123  &      77  &      96  &     29.0s  &      6.5s  &      5.4s  &     11.9s  \\
\d{paperView}                   &        63  &      61  &      85  &      8.0s  &      0.8s  &      2.0s  &      2.8s  \\
\d{reviewsView}                 &        96  &      61  &      70  &     14.6s  &      5.0s  &      0.8s  &      5.8s  \\
\d{profileViewGet}              &        46  &       0  &       0  &      0.2s  &      0.2s  &      0.0s  &      0.2s  \\
\d{profileViewPost}             &        20  &       0  &       0  &      0.1s  &      0.1s  &      0.0s  &      0.1s  \\
\d{submitReviewViewPost}        &       103  &       0  &       0  &      9.0s  &      7.8s  &      0.0s  &      7.8s  \\
\d{assignReviewersView}         &        63  &       0  &       0  &      0.6s  &      0.6s  &      0.0s  &      0.6s  \\
\midrule
Total\st                       &       627  &     215  &     276  &     64.1s  &     23.5s  &      8.4s  &     31.9s  \\
\bottomrule
\end{tabular}
}

\bigskip
  \noindent \hspace{1cm} {\sffamily\footnotesize (b) Gradr---Course Management System}\hfill 
  \hbox{\scalebox{0.80}{\fbox{Policy size (AST nodes):  75}} \hspace{1cm}} 
\\
\renewcommand\d[2]{\lookupPut{#1}{#2}}
\d{homePage}{Display the home page (static content)}
\d{profileView}{View a user's profile (owner)}
\d{unauthProfileView}{View a user's profile (any user)}
\d{scoresForAssignmentView}{Instructor: view scores for an assignment}
\d{scoresForStudentView}{Student: view all scores for user}
\d{topScoreForAssignmentView}{Instructor: view top scores for an assignment}
\renewcommand\d[1]{\lookupGet{#1}} 
\renewcommand\arraystretch{1}
\scalebox{0.80}{
\begin{tabular}{l|rr|rrr}
\toprule
  \multirow{2}{*}{\textbf{Benchmark}} & \multicolumn{2}{c|}{\textbf{Code size (AST nodes)}} & \multicolumn{3}{c}{\textbf{Time}} \\
&  Original  &   \toolname     &  Localize            &     Generate  &  Total  \\
\midrule
\d{homePage}                              &        17  &       0  &      0.0s  &      0.0s  &      0.0s  \\
\d{profileView}                           &       119  &      59  &      3.0s  &      0.8s  &      3.8s  \\
\d{unauthProfileView}                     &       120  &      59  &      3.1s  &      1.5s  &      4.6s  \\
\d{scoresForAssignmentView}               &        73  &     103  &      0.9s  &      1.4s  &      2.3s  \\
\d{topScoreForAssignmentView}             &        99  &     147  &      2.4s  &      2.4s  &      4.8s  \\
\d{scoresForStudentView}                  &       100  &     112  &      3.7s  &      9.5s  &     13.2s  \\
\midrule
Total\st                                 &       528  &     480  &     13.2s  &     15.6s  &     28.8s  \\
\bottomrule
\end{tabular}
}

\bigskip
\noindent \hspace{1cm} {\sffamily\footnotesize (c) HealthWeb---Health Information Portal} \hfill 
  \hbox{\scalebox{0.80}{\fbox{Policy size (AST nodes):  95}} \hspace{1cm}} 
\\
\renewcommand\d[2]{\lookupPut{#1}{#2}}
\d{showRecordByIdView}{Search a record by id}
\d{showRecordsForPatientView}{Search a record by patient}
\d{showAuthoredRecordsView}{Show authored records}
\d{updateRecordForm}{Update record}
\d{listOfPatientsView}{List patients for a doctor}
\renewcommand\d[1]{\lookupGet{#1}} 
\scalebox{0.80}{
\begin{tabular}{l|rr|rrr}
\toprule
\multirow{2}{*}{\textbf{Benchmark}} & \multicolumn{2}{c|}{\textbf{Code size (AST nodes)}} & \multicolumn{3}{c}{\textbf{Time}} \\
&  Original  &   \toolname     &  Localize            &     Generate  &  Total  \\
\midrule
\d{showRecordByIdView}                    &        27  &     161  &      0.1s  &     13.0s  &     13.1s  \\
\d{showRecordsForPatientView}             &        74  &     339  &      1.2s  &     41.6s  &     42.8s  \\
\d{showAuthoredRecordsView}               &        76  &       0  &      1.2s  &      0.0s  &      1.2s  \\
\d{updateRecordForm}                      &        25  &       0  &      0.0s  &      0.0s  &      0.0s  \\
\d{listOfPatientsView}                    &        80  &      87  &      1.3s  &      3.5s  &      4.8s  \\
\midrule
Total\st                                 &       282  &     587  &      3.8s  &     58.1s  &     61.9s  \\
\bottomrule
\end{tabular}
}
\caption{\label{eval:benchmarks:case studies}
  Case studies: conference management, course manager, health portal.}
\vspace{-.5cm}  
\end{table*}

\custompar{Implementation.}
We have implemented a prototype \toolname compiler
by extending the \synquid program synthesizer~\cite{PolikarpovaKuSo16}.
From \synquid, \toolname inherits a liquid type checker and a type-driven synthesis mechanism.
On top of this, our implementation adds
\begin{inparaenum}[(1)]
\item the \T{TIO} library, which implements the API shown in \autoref{code:primitives} 
plus some standard output actions and redaction functions (90 lines of \toolname code);
\item the implementation of the \textproc{Enforce} algorithm from \autoref{sec:synthesis} that calls out to the type-checker and the synthesizer;
\item a \synquid-to-Haskell translator, which can link \toolname code with other Haskell modules.
\end{inparaenum}
Thanks to the translator, 
a possible usage scenario for \toolname is to serve as a language for the data-centric application core,
while low-level libraries can be implemented directly in Haskell.

\custompar{Programs.}
To evaluate the \toolname compiler, we implemented 
\begin{inparaenum}[(1)]
\item a set of microbenchmarks that highlight challenging scenarios and model reported real-world leaks; and
\item three larger case studies based on existing applications from the literature.
\end{inparaenum}
For each of these programs, we specified the security policies in the style of \autoref{code:edas-policies}
and implemented the basic logic of the controllers \emph{omitting all policy checks}.
Hence, for each controller, \toolname must localize unsafe data accesses 
and generate leak patches.
For one of our case studies, we additionally implemented a non-leaky version with manually written policy checks.

\custompar{Evaluation criteria.}
Our goal is to evaluate the following parameters:
\begin{itemize}
\item\textbf{Expressiveness of policy language.} 
We demonstrate that \toolname is expressive enough to support interesting policies from in a range of problem domains,
including conference management, course management, health records, and social networks.
In particular, we were able to replicate all the desired policies in three case studies from prior work~\cite{SwamyCC10,Yang2016}.

\item\textbf{Performance.} We show that the \toolname compiler is reasonably efficient at leak localization and patch synthesis:
\toolname is able to generate all necessary patches for each of our case studies
in 30--60 seconds.

\item\textbf{Quality of patches.}
We compare the code generated by \toolname to a version with manual policy enforcement
and show that it is able to recover \emph{all} necessary policy-enforcing code, without reducing functionality.
\end{itemize}


\subsection{Microbenchmarks}

To exercise the flexibility of our language,
we implemented a series of small but challenging microbenchmarks,
summarized in \autoref{eval:benchmarks:micro}.
The code of each benchmark, with leak patches inserted by \toolname,
is available in \autoref{appendix:benchmarks}. 

Benchmark 1 is our running example from \autoref{sec:example};
benchmarks 2--3 are its variations with multiple unsafe accesses in the same controller
and with a self-referential policy on the ``authors'' field, respectively.
Benchmarks 4 and 5 exercise tricky cases of implicit flow through higher-order functions.
\T{Search} is the controller from \autoref{code:search-leak},
which displays the titles of all \T{client}'s accepted papers;
here \toolname inserts a patch inside the \T{filterM}'s predicate.

\T{Sort} displays the list of all conference submissions sorted by their score, 
using a higher-order \T{sortM} function with a custom comparator.
The order of submissions might leak paper scores to a conflicted reviewer. 
To prevent this leak, \toolname rewrites the comparator to return a default score 
if a paper is conflicted with the viewer. 
Interestingly, this benchmark features a \emph{negative self-referential policy} for the list of conflicted reviewers:
this list is visible only to users who are \emph{not} on the list.
Such policies are not supported by Jeeves,
since they are incompatible with its fixpoint interpretation of self-referential policies (\autoref{sec:related});
in \toolname, the semantics of policies is decoupled from their evaluation,
hence this example presents no difficulty.

\T{Broadcast} sends a decision notification to all authors of a given paper.
This benchmark tests \toolname's ability to handle messages sent to multiple users;
\toolname infers that all those users are authors of the paper, and hence are allowed to see its decision,
as long as the phase is \T{Done}.
An additional challenge is that the list of recipients is \emph{itself sensitive}, since the conference is double-blind;
\toolname infers that no additional check is needed, since authors are always allowed to see themselves.

The last three benchmarks model reported real-world leaks.
\T{HotCRP} models a leak in the HotCRP conference manager, first reported in~\cite{Yip2009},
where the conference chair could send password reminder emails to PC members,
and then glean their passwords from the email preview.
\toolname repairs this leak by masking the password in the preview (but not in the actual email),
since the preview is flowing to the chair, while the email is flowing to the owner of the password.

\T{AirBnB} models a leak in the AirBnB website~\cite{AirBNBexample:private}.
The website redacts phone numbers from user messages
(presumably to keep people from going around the site), 
but phone numbers appear unredacted in message previews.
In \toolname we model the AirBnB messaging system by designating the message text visible only to its sender and the site administrator,
and introducing a special redaction function \T{scrubPhoneNumbers},
whose result is additionally visible to the message recipient.
With these policies, whenever a message is displayed to the recipient,
\toolname inserts a check whether they are the administrator, 
and otherwise redacts the text with \T{scrubPhoneNumbers}.

\T{Instagram} is inspired by several reported cases, 
where sensitive social network data was revealed through recommendation algorithms~\cite{Comey,SexWorkers}.
In particular, if an Instagram account is private,
their photos and ``following'' relations are supposedly only visible to their followers
(which have to be approved by the user).
Yet, journalist Ashley Feinberg was able to identify the private Instagram account of the former FBI director James Comey,
because Instagram mistakenly revealed that James was followed by his son Brien (whose account is public).
In \toolname, we model the Instagram ``following'' relation using a getter, whose policy requires that both accounts be visible to the viewer:
\begin{lifty}[numbers=none]
measure following :: Store -> User -> Set User
getIsFollowing :: ds: Store -> who: User -> whom: User ->
  TI {Bool|_v == (whom in following ds who)} <canSee ds _u who && canSee ds _u whom>
inline canSee ds x y  =  x == y || isPublic ds y || y _in_ following ds x
\end{lifty}
When the recommendation system attempts to retrieve all accounts followed by Brien Comey,
\toolname injects a check that those accounts be visible to the viewer,
and otherwise replaces the true value of ``is following'' with false.

\subsection{Case Studies}

We use \toolname to implement three larger case studies:
a conference manager and a course manager, both based on examples from Jacqueline~\cite{Yang2016},
and a health portal based on the HealthWeb example from Fine~\cite{SwamyCC10}.
\autoref{eval:benchmarks:case studies} lists the controllers we implemented for each case study,
together with their sizes in AST nodes.
More precisely, the column ``Original'' refers to the size of manually-written code without policy checks,
and ``\toolname'' refers to the size of auto-generated policy-enforcing code;
for an example, see grayed-out vs highlighted code in \autoref{code:edas-fixed}.
Conference manager additionally reports the size of manually written policy-enforcing code in the column ``Manual''.
For each case study, we also report the size of the policy, which is the cumulative size of all refinements of input and output actions,
plus the size of \T{inline} macro definitions;
we do not include the size of the \toolname standard library into the policy size.

\custompar{Conference manager.}
We implemented two versions of a basic academic conference manager: 
one where the programmer enforces the policies by hand
(and \toolname only verifiers correctness)
and one where the programmer omits all policy-enforcing code
(and \toolname is responsible for injecting leak patches).
The former version contains 247 lines of \toolname code
while the latter contains 216;
both systems share 364 lines of Haskell code that implement non-security-critical functionality.
The manager handles confidentiality policies for user profiles, submissions, and reviews,
and enforces policies such as:
``a user profile is only visible to that user and the conference chair'' or
``the list of PC members conflicted with a submission is only visible to PC members who are not conflicted''.
%

While Jacqueline only supports constant default values,
we decided to deviate from the original system to experiment with nontrivial redaction functions.
In our version, reviewer names that are hidden for any reason are displayed as ``Reviewer A'', ``Reviewer B'', etc.,
following common convention. 
This is implemented by representing each reviewer entry
as a pair of (\textit{index}, \textit{name}), where the redaction replaces \textit{name}
with ``Reviewer \textit{x}\hspace{1pt}'' according to \textit{index}.

\custompar{Course manager.} We implemented a system for sending grades to students 
based on their course enrollment and assignment status.
An example policy is that a student can see their own scores, whereas instructors can see scores for all of their students.

\custompar{Health portal.} Based on the HealthWeb case study from~\cite{SwamyCC10}, 
we implemented a system that supports viewing and searching over health records.
This case study is interesting because of the complexity of the associated policies.
For example, the policy that guards patients associated with health records
states that the viewer must be the author of the record or the patient;
otherwise non-withheld records can be viewed by a doctor, 
but psychiatric records can only be viewed by the doctor actually treating the patient:
\begin{lifty}[numbers=none]
inline recordPolicy w v rid = v == recordAuthor w rid || v == recordPatient w rid || 
 (!(shouldWithhold w rid) && isDoctor w v && 
 (isTreating w v (recordPatient w rid) || !(recordIsPsychiatric w rid)))   
getRecordPatient :: ds: Store -> rid: RecordId 
  -> TI {User | _v == recordPatient ds rid} <recordPolicy ds _u rid>
\end{lifty}
As a result, the generated policy enforcement code for this study 
is significantly larger than the original program,
and takes twice as long to generate as in the conference manager.




\subsection{Performance Statistics}
We show compilation times for the microbenchmarks in \autorefs{eval:benchmarks:micro}, 
and for the case studies in \autorefs{eval:benchmarks:case studies}.
We break them down into leak localization (including type checking) and patch synthesis.
\toolname was able to patch each of the microbenchmarks in under two second.
%
%
For each of the three case studies, \toolname takes 30--60 seconds.
Interestingly, the version of the conference manager with manual policy enforcement
takes \emph{longer} to verify than the leaky version takes to repair (64 vs 32 seconds).
This is a side-effect of the restriction imposed by our repair algorithm (\autoref{sec:synthesis:localization}) 
that expected types for patches be functionally oblivious.
Thanks to this restriction, automatically-generated leak patches can be verified
independently from the rest of the controller (and from each other);
on the other hand, with manual policy enforcement
\toolname verifies the controller function as a whole,
which is less efficient but more precise.
%

\custompar{Scalability.}
Note that \toolname verifies and patches each top-level function in a program completely independently.
Moreover, unlike prior work on program repair,
patch synthesis proceeds independently for different leaks inside one function,
which allows \toolname to scale to functions that require multiple patches.
For a stress test, we created a benchmark that sequences together $N$ reads of a sensitive field,
and then a \T{print} to an arbitrary user.
\toolname's job is to patch each of the $N$ leaks.
Both leak localization and patch generation scale non-linearly but relatively well:
as $N$ grows from 1 to 16,
the total compilation time increases from less than a second to just over $20$ seconds.
The repair time is non-linear because with the sequential structure of our benchmark each read introduces a new variable,
which is visible in all the following patches,
and hence increases the search space for policy checks.
You can find the detailed results of this experiment is \autoref{appendix:scalability}.



\subsection{Quality of Patches}

We compared the two versions of our conference manager (\autoref{eval:benchmarks:case studies}).
The column ``Original'' shows the size of the code, in AST nodes, without any policy enforcement.
We also show the cumulative size of policy-enforcing code, 
both hand-written and generated by \toolname.
Note that the size of policy-enforcing code often approaches or exceeds the size of the core functionality, 
which motivates the \toolname repair engine as an approach to reducing the programmer burden.
Manual inspection reveals that while the two versions of policy-enforcing code are syntactically different,
they differ in neither functionality nor performance.

\subsection{Discussion and Limitations}

We conclude this section with a discussion of \toolname's limitations.

\custompar{Policy Language.}
The expressiveness of \toolname policies is limited by the underlying SMT theory of quantifier-free linear arithmetic, uninterpreted functions, and arrays.
For example, policies cannot involve non-linear arithmetic or arbitrary recursive functions over data 
(e.g., refer to the maximum of all paper scores).
Since all functions are uninterpreted, \toolname does not automatically know, for example, 
that paper reviewers cannot be in conflict with a paper. 
This could result in generating redundant checks but is also easy to avoid by
adding postconditions to the input action \T{getReviewers}, 
relating it not only to the \T{reviewers} measure but to the \T{conflicts} measure.
Perhaps most importantly, complex policies are most naturally expressed using existential quantification;
for example, to state that a review $r$ is visible only to reviewers that have submitted a reviews for the same paper,
we would like to write: $\exists r' . \T{reviewPaper}~r = \T{reviewPaper}~r' \wedge \upsilon = \T{reviewer}~r'$.
This policy is currently not supported by \toolname: 
instead, a programmer would need to introduce ``inverse measures'' for \T{reviewPaper} and \T{reviewer}
and write $\T{reviewsBy}~\upsilon \cap \T{reviewsFor}~(\T{reviewPaper}~r) \neq \emptyset$
(and add postconditions to getters to connect direct and inverse measure).

\custompar{Programming Model.}
While the present work lays a foundation for static IFC with liquid types,
our ultimate goal is to turn \toolname into a realistic web framework for Haskell,
building upon \LH~\cite{VazouSeJh14}.
Apart from a significant engineering effort,
there are several research challenges involved in achieving this goal: 
most importantly, encoding a realistic database model 
that supports provably secure interaction with the data store through SQL-like \emph{queries} rather than retrieving fields ``one at a time''.
This is challenging because the type checker has to infer an aggregate label for all the data returned by the query,
which has to be precise enough to verify common data retrieval patterns.
In addition, we will need a more convenient way to specify existential policies mentioned above,
a way to generate input-output actions automatically from a declarative description of the application's \emph{model} 
(\ie database schema with policies),
as well as a way to tie the \T{TIO} monad into a server framework.
We leave all these improvements to future work.

\section{Related Work}\label{sec:related}

\toolname builds upon several lines of prior work, most notably in static
information flow control, program synthesis and repair, and type error
localization.  Each of these areas has a rich history, but until now they
have developed relatively independently.

\subsection{Information Flow Control}

The \toolname type system builds upon a long history of work in language-based
information flow control~\cite{Sabelfeld2003}.
Though we (indirectly) borrow some ideas from dynamic IFC systems---in
particular Hails/LIO~\cite{giffin:2017:hails, StefanM14, StefanMMR17} and
Jeeves~\cite{Yang2012, Austin2013, Yang2016}---\toolname enforces security
policies using an information-flow type system.
We see our work as complimentary to previous efforts on static information flow
type systems.
For example, Jif~\cite{Jif0}, Fabric~\cite{FabricSOSP,OaklandFabric} and
Paragon~\cite{paragon} have been used to enforce IFC for Java programs,
FlowCaml~\cite{flowcaml} for OCaml, and SLIO~\cite{buiras2015hlio},
MAC~\cite{vassena2018mac}, and
others~\cite{LiZ06,Russo15,hughes00generalising,Russo08,Devriese:2011}
for Haskell.
To our knowledge, \toolname is the first system to encode IFC into the
framework of liquid types---and while our implementation is for \Synquid, we
think \toolname can be similarly be implemented in other languages with liquid
types (\eg \LH~\cite{VazouSeJh14}).

Our IFC encoding shares some similarities to SLIO~\cite{buiras2015hlio},
Fine~\cite{SwamyCC10, FineCompiler} and F$^*$~\cite{FStar}.
Like \toolname, all three use a monadic encoding of information flow; many
others share a similar encoding~\cite{LiZ06, Russo08, Russo15, VassenaR16, Crary:2005},
going back as far as the Dependency Core Calculus~\cite{dcc}.
Fine~\cite{SwamyCC10, FineCompiler}, F$^*$~\cite{FStar}, and
others~\cite{lourencco2014information, lourencco2015dependent} additionally
support value-dependent security types.
The key difference is that our system uses (SMT-decidable) predicates as security labels,
which supports
\begin{inparaenum}[(1)]
\item a direct encoding of Hails- and Jeeves-like policies, and
\item fully automatic verification and leak localization, crucial for repair. 
\end{inparaenum}
UrFlow~\cite{Chlipala10} is the only automated verification system that supports a similar flavor of policies,
but it does not provide a sound treatment of self-referential policies.
More importantly, none of these approaches address the issue of programmer burden: 
they simply prevent unsafe programs from compiling, but do not help programmers write policy-enforcing code.

Our policies are closer to policies used in IFC web frameworks (e.g.,
Hails~\cite{giffin:2017:hails} and Jacqueline~\cite{Yang2016}) than most IFC
systems.
Indeed, most IFC systems track the flow of information by associating labels
with data and thus need to keep labels simple to be efficient.
While existing label models can be used to encode web application
policies~\cite{dlm, dclabels, GenLabels}, high-level declarative policies like
\toolname's are usually instantiated to these labels.
Jeeves~\cite{Yang2012} and Nexus~\cite{nexus} are exceptions to these---they
respectively encode policies in SMT-decidable and first-order logics---but enforce
these policies at run time.
Beyond runtime overhead, such rich policies are also harder to debug at
runtime---\eg in Jeeves this is the case because the runtime replaces sensitive
values with default values when the policy is not satisfied.

\subsection{Program Synthesis and Repair}

\toolname is related to techniques for synthesizing provably correct programs from formal specifications~\cite{MannaWa80,KuncakMaPiSu10,KneussKKS13,PolikarpovaKuSo16,AlurRU17}.
These techniques, however, generate programs from scratch, from end-to-end functional specifications,
while \toolname injects code into an existing program based on the cross-cutting concern of information flow.

Our problem statement is similar to that of deductive program repair~\cite{KneussKK15},
but in the specific setting of policy enforcement 
\toolname is able to infer a local specification for each patch,
and synthesize all patches independently,
which makes it more scalable.
There has been prior work on program repair for security concerns~\cite{Harris2010,FredriksonJJRPSY12,SonMS13,Ganapathy2006},
but it does not involve reasoning about expressive information-flow policies,
and hence, both the search space for patches and their verification is much less complex.
Finally, our repair technique is based on CHCs and uses a Horn solver to infer the optimal expected type;
\cite{HojjatRMCF16} also propose a Horn-based repair technique but for a different domain (software-defined networks).

\subsection{Type Coercions and Type Error Localization}

Our use of type errors to target program rewriting resembles \emph{type-directed coercion insertion}~\cite{Swamy2009TTC};
in particular, their coercion insertion and coercion generation mechanisms 
are similar to our fault localization and patch synthesis, respectively,
and their coercion set is similar to our set $\mathcal{R}$ of redaction functions.
The \toolname type system, however, is far more expressive than the type systems explored in that work.
In particular, the combination of polymorphism and subtyping complicates type error localization (since there are many valid type derivations),
while refinements complicate coercion generation (which becomes a refinement type inhabitation problem).

Hybrid type checking~\cite{KnowlesF10} can be viewed as coercion insertion for refinement types.
In fact, their coercions also amount to wrapping the original value in a conditional,
however, in their case both the guard and the alternative branch are straightforward.

Existing work on type error localization for expressive types systems~\cite{zmvp15,Loncaric2016,SeidelSCWJ17}
is in a more general---yet more forgiving---context of giving feedback to programmers.
Our leak localization technique (removing constraints that make the system unsatisfiable) is similar to~\cite{Loncaric2016},
but for our specific purpose we have more information to decide between possible error locations.

\begin{acks}
  The authors would like to thank the anonymous reviewers
  and our shepherd, Nikhil Swamy, for their valuable feedback on earlier drafts of this paper.
  We are also grateful to Marco Vassena for suggesting
  how to simplify and strengthen the noninterference theorem.
  This work was supported by a gift from Cisco and by the National Science Foundation under Grants No.~1911149 and 1943623.

\end{acks}


\clearpage
\appendix
\section{The Language \corelg}\label{appendix:language}

\subsection{Operational Semantics of \corelg}

The full operational semantics of \corelg is given in \autoref{formalism:operational-full}.

\begin{figure*}
\addtolength{\jot}{2mm}
\textbf{Evaluation}\quad$\boxed{\config{\store}{t} \bigeval \config{\store'}{v}}$
\smallskip
\begin{gather*}
  \inference[\textsc{ret}]
    {}
    {\config{\store}{\kwreturn~t} \bigeval \config{\store}{\tioval{t}}}
  \quad
  \inference[\textsc{bind}]
    {\config{\store}{t_1} \bigeval \config{\store'}{\tioval{t_1'}} &
      \config{\store'}{t_2 ~ t_1'} \bigeval \config{\store''}{v}}
    {\config{\store}{\kwbind ~ t_1 ~ t_2} \bigeval \config{\store''}{v}}      
  \\
  \inference[\textsc{down}]
    {\config{\store}{t_1} \bigeval \config{\store}{f} & \config{\store}{t_2} \bigeval \config{\store'}{\tioval{t}} & \config{\store'}{t} \bigeval \config{\store'}{b}}
    {\config{\store}{\kwdown~t_1~t_2} \bigeval \config{\store'}{\tioval{b}}}  
  \\
  \inference[\textsc{get}]
    {\config{\store}{t} \bigeval \config{\store}{f}}
    {\config{\store}{\kwget ~ t} \bigeval \config{\store}{\tioval{\store[f]}}}
  \quad
  \inference[\textsc{set}]
    {\config{\store}{t_1} \bigeval \config{\store}{f} & \config{\store}{t_2} \bigeval \config{\store}{b}}
    {\config{\store}{\kwset ~ t_1 ~ t_2} \bigeval \config{\store[f := b]}{\tioval{\unit}}}  
  \\
  \inference[\textsc{if-true}]
    {\config{\store}{t} \bigeval \config{\store}{\vtrue} & \config{\store}{t_1}\bigeval\config{\store'}{v}}
    {\config{\store}{\kwif ~ t ~ \kwthen ~ t_1 ~ \kwelse ~ t_2} \bigeval \config{\store'}{v}}
  \\
  \inference[\textsc{if-false}]
    {\config{\store}{t} \bigeval \config{\store}{\vfalse} & \config{\store}{t_2}\bigeval\config{\store'}{v}}
    {\config{\store}{\kwif ~ t ~ \kwthen ~ t_1 ~ \kwelse ~ t_2} \bigeval \config{\store'}{v}}
  \\
  \inference[\textsc{app}]
    {\config{\store}{t_1} \bigeval \config{\store}{\lambda x.~ t_1'} & \config{\store}{[x\,{\mapsto}\,t_2]t_1'} \bigeval \config{\store'}{v}}
    {\config{\store}{t_1 ~ t_2} \bigeval \config{\store'}{v}}
  \quad  
  \inference[\textsc{val}]
    {}
    {\config{\store}{v} \bigeval \config{\store}{v}}
\end{gather*}
\caption{\label{formalism:operational-full}
  Big-step operational semantics of \corelg.}
\end{figure*}

\subsection{The \corelg Type System}\label{appendix:types}

\begin{figure*}
\addtolength{\jot}{2mm}
\renewcommand\colon{{\,:\,}}  
\textbf{Well-formedness}\quad$\boxed{\env \vdash T}$
\begin{gather*}
  \inference[\textsc{wf-base}]
    {\env, \nu\colon B \vdash r : \T{Bool}}
    {\env \vdash \{B \mid r\}}
  \quad
  \inference[\textsc{wf-fun}]
    {\env \vdash T_1 & \env \vdash T_2}
    {\env \vdash \funT{x}{T_1}{T_2}}  
  \quad
  \inference[\textsc{wf-Field}]
    {\env\vdash \{B \mid r\} & \env,\upsilon \colon\T{User} \vdash l: \T{Bool}}
    {\env \vdash \fld{\{B \mid r\}}{l} }  
  \\
  \inference[\textsc{wf-TIO}]
    {\env\vdash T & \env,\upsilon \colon\T{User} \vdash l_i: \T{Bool} & \env,\upsilon\colon\T{User} \vdash l_o: \T{Bool}}
    {\env \vdash \tio{T}{l_i}{l_o} }
\end{gather*}    

\medskip
\textbf{Subtyping and Flow}\quad$\boxed{\env \vdash T \Subt T'}$\quad$\boxed{\env \vdash l \Flows l'}$
\begin{gather*}
  \inference[\textsc{$\Subt$-Base}]
    {\env \vDash r\Implies r'}
    {\env \vdash \{B\mid r\} \;\Subt\; \{B\mid r'\} }
  \quad
  \inference[\textsc{$\Subt$-Fun}]
    {\env\vdash T_1' ~\Subt~ T_1 & \env \vdash T_2 ~\Subt~ T_2'}
    {\env \vdash \funT{x}{T_1}{T_2} ~\Subt~ \funT{x}{T_1'}{T_2'} }
  \\
  \inference[\textsc{$\Subt$-TIO}]
    {\env \vdash T_1 \Subt T_2 & \env \vdash l_1 \Flows l_2 & \env \vdash l_2' \Flows l_1'}
    {\env \vdash \tio{T_1}{l_1}{l_1'} \Subt \tio{T_2}{l_2}{l_2'}}
  \quad
  \inference[\textsc{flow}]
    {\env, \upsilon\colon\T{User} \vDash l' \Implies l}
    {\env \vdash l \Flows l'}
  \\
  \inference[\textsc{$\Subt$-Field}]
    { }
    {\env \vdash \fld{T}{l} \Subt \fld{T}{l}}    
\end{gather*}

\medskip
\textbf{Typing}\quad$\boxed{\env \vdash t::T}$
\begin{gather*}
  \inference[\textsc{t-True}]
    {}
    {\env \vdash \vtrue :: \{\T{Bool} \mid \nu\}}
  \quad
  \inference[\textsc{t-False}]
    {}
    {\env \vdash \vfalse :: \{\T{Bool} \mid \neg\nu\}}
  \quad
  \inference[\textsc{t-Unit}]
    {}
    {\env \vdash \unit :: \T{()}}
  \\
  \inference[\textsc{t-User}]
    {}
    {\env \vdash \vuser_i :: \T{User}}
  \quad    
  \inference[\textsc{t-$\lambda$}]
    {\env, x:T_1 \vdash t::T_2}
    {\env \vdash \lambda x.~t :: T_1\to T_2}
  \quad
  \inference[\textsc{t-Field}]
    {\typeof(\vfield_i) = T & \labelof(\vfield_i) = l}
    {\env \vdash \vfield_i :: \fld{T}{l}}
  \\
  \inference[\textsc{t-TIO}]
    {\env \vdash t::T}
    {\env \vdash \tioval{t} :: \tio{T}{\bot}{\top}}
  \quad
  \inference[\textsc{t-var}]
    {x\colon T \in \env}
    {\env \vdash x :: T}
  \quad
  \inference[\textsc{t-app}]
    {\env \vdash t_1 :: \funT{x}{T}{T'} & \env \vdash t_2::T}
    {\env \vdash t_1 ~ t_2 :: T'}
  \\
  \inference[\textsc{t-if}]
    {\env \vdash t :: \{\T{Bool} \mid r\} \\
     \env, [\nu\mapsto\true]r \vdash t_1::T  &  \env, [\nu\mapsto\false]r \vdash t_2::T}
    {\env \vdash \kwif ~ t ~ \kwthen ~ t_1 ~ \kwelse ~ t_2 :: T} \quad
  \\
  \inference[\textsc{t-ret}]
    {\env \vdash t :: T}
    {\env \vdash \kwreturn ~ t :: \tio{T}{\bot}{\top}}
  \\  
  \inference[\textsc{t-bind}]
    {\env \vdash t_1 :: \tio{T_1}{l_1}{l_1'} & \env \vdash t_2 :: \funT{x}{T_1}{\tio{T_2}{l_2}{l_2'}} & \env \vdash l_1 \Flows l_2'}
    {\env \vdash \kwbind ~ t_1 ~ t_2 :: \tio{T_2}{l_1 \Fjoin l_2}{l_1' \Fmeet l_2'}}
  \\
  \inference[\textsc{t-down}]
    {\env \vdash t_1 :: \fld{\_}{l} & \env \vdash t_2 :: \tio{\{\T{Bool} \mid \nu \Implies r\}}{l \Fjoin r}{l'}}
    {\env \vdash \kwdown ~ t_1 ~ t_2 :: \tio{\T{Bool}}{l}{l'}}
  \\
  \inference[\textsc{t-get}]
    {\env \vdash t :: \fld{T}{l}}
    {\env \vdash \kwget ~ t :: \tio{T}{l}{\top}}
  \quad
  \inference[\textsc{t-set}]
    {\env \vdash t_1 :: \fld{T}{l} & \env \vdash t_2 :: T}    
    {\env \vdash \kwset ~ t_1 ~ t_2 :: \tio{\T{()}}{\bot}{l}}
  \\
  \inference[\textsc{t-$\Subt$}]
    {\env \vdash t :: T' & \env \vdash T'\Subt T}
    {\env \vdash t :: T}
\end{gather*}    

\caption{\label{formalism:types-full}
  \corelg static semantics.}
\end{figure*}

The full static semantics of \corelg is given in \autoref{formalism:types-full}.

\begin{figure*}
\addtolength{\jot}{2mm}
\textbf{Instrumented Evaluation}\quad$\boxed{\config{\store,\instr{\ell_c}}{t} \liobigeval \config{\store',\instr{\ell'_c}}{v}}$
\smallskip
\begin{gather*}
  \inference[\textsc{i-ret}]
    {}
    {\config{\store,\instr{\ell_c}}{\kwreturn~t} \liobigeval \config{\store,\instr{\ell_c}}{\tioval{t}}}
  \\
  \inference[\textsc{i-bind}]
    {\config{\store,\instr{\ell_c}}{t_1} \liobigeval \config{\store',\instr{\ell'_c}}{\tioval{t_1'}} &
    \config{\store',\instr{\ell'_c}}{t_2 ~ t_1'} \liobigeval \config{\store'',\instr{\ell''_c}}{v}}
    {\config{\store,\instr{\ell_c}}{\kwbind ~ t_1 ~ t_2} \liobigeval \config{\store'',\instr{\ell''_c}}{v}}    
  \\
  \inference[\textsc{i-down-1}]
    {\config{\store,\instr{\ell_c}}{t_1} \liobigeval \config{\store,\instr{\ell_c}}{f} &
     \config{\store,\instr{\ell_c}}{t_2} \liobigeval \config{\store',\instr{\ell'_c}}{\tioval{t}} &      
     \instr{\sem{\labelof(f)} = \ell} &
     \instr{\ell'_c \Flows \ell \Fjoin \ell_c} \\
     \config{\store',\instr{\ell'_c}}{t} \liobigeval \config{\store',\instr{\ell'_c}}{b}
     }
    {\config{\store,\instr{\ell_c}}{\kwdown~t_1~t_2} \liobigeval \config{\store',\instr{\ell\Fjoin\ell_c}}{\tioval{b}}}
  \\
  \inference[\textsc{i-down-2}]    
    {\config{\store,\instr{\ell_c}}{t_1} \liobigeval \config{\store,\instr{\ell_c}}{f} &
     \config{\store,\instr{\ell_c}}{t_2} \liobigeval \config{\store',\instr{\ell'_c}}{\tioval{t}} & 
     \instr{\sem{\labelof(f)} = \ell} &
     \instr{\ell'_c \not\Flows \ell \Fjoin \ell_c} \\
     \config{\store',\instr{\ell'_c}}{t} \liobigeval \config{\store',\instr{\ell'_c}}{b}
    }
    {\config{\store,\instr{\ell_c}}{\kwdown~t_1~t_2} \liobigeval \config{\store',\instr{\ell\Fjoin\ell_c}}{\tioval{\vfalse}}}
  \\
  \inference[\textsc{i-get}]
    {\config{\store,\instr{\ell_c}}{t} \bigeval \config{\store,\instr{\ell_c}}{f} & \instr{\sem{\labelof(f)} = \ell}}
    {\config{\store,\instr{\ell_c}}{\kwget ~ t} \liobigeval \config{\store,\instr{\ell\Fjoin \ell_c}}{\tioval{\store[f]}}}
  \\
  \inference[\textsc{i-set}]
    {\config{\store,\instr{\ell_c}}{t_1} \bigeval \config{\store,\instr{\ell_c}}{f} &
     \config{\store,\instr{\ell_c}}{t_2} \liobigeval \config{\store,\instr{\ell_c}}{b} & 
     \instr{\ell_c \Flows \sem{\labelof(f)}}}
    {\config{\store,\instr{\ell_c}}{\kwset ~ t_1 ~ t_2} \liobigeval \config{\store[f := b],\instr{\ell_c}}{\tioval{\unit}}}
  \\
  \inference[\textsc{i-if-true}]
    {\config{\store,\instr{\ell_c}}{t} \liobigeval \config{\store,\instr{\ell_c}}{\vtrue} & 
     \config{\store,\instr{\ell_c}}{t_1}\liobigeval\config{\store',\instr{\ell'_c}}{v}}
    {\config{\store,\instr{\ell_c}}{\kwif ~ t ~ \kwthen ~ t_1 ~ \kwelse ~ t_2} \liobigeval \config{\store',\instr{\ell'_c}}{v}}
  \\
  \inference[\textsc{i-if-false}]
    {\config{\store,\instr{\ell_c}}{t} \liobigeval \config{\store,\instr{\ell_c}}{\vfalse} & 
     \config{\store,\instr{\ell_c}}{t_2}\liobigeval\config{\store',\instr{\ell'_c}}{v}}
    {\config{\store,\instr{\ell_c}}{\kwif ~ t ~ \kwthen ~ t_1 ~ \kwelse ~ t_2} \liobigeval \config{\store',\instr{\ell'_c}}{v}}
  \\
  \inference[\textsc{i-app}]
    {\config{\store,\instr{\ell_c}}{t_1} \liobigeval \config{\store,\instr{\ell_c}}{\lambda x.~ t_1'} & 
     \config{\store,\instr{\ell_c}}{[x\,{\mapsto}\,t_2]t_1'} \liobigeval \config{\store',\instr{\ell'_c}}{v}}
    {\config{\store,\instr{\ell_c}}{t_1 ~ t_2} \liobigeval \config{\store',\instr{\ell'_c}}{v}}    
  \\
  \inference[\textsc{i-val}]
    {}
    {\config{\store,\instr{\ell_c}}{v} \bigeval \config{\store,\instr{\ell_c}}{v}}    
\end{gather*}
\caption{\label{formalism:instrumented-full}
  Instrumented operational semantics.}
\end{figure*}

\subsection{Noninterference}

\custompar{Instrumented Semantics.}
The full instrumented operational semantics is given in \autoref{formalism:instrumented-full}.

\custompar{From Instrumented Semantics to \LIO.}
To formalize the noninterference guarantee, we extend our calculus with erased terms, denoted with \hole.
We define the \emph{erasure function} on stores and instrumented configurations as follows:
\begin{align*}
\erase{\ell}{\store}[f] &= 
  \begin{cases}
  \store[f] \quad \sem{\labelof(f)}\Flows \ell\\
  \hole \quad\text{otherwise}
  \end{cases}\\
\erase{\ell}{\config{\store, \ell_c}{t}} &= 
  \begin{cases}
  \config{\erase{\ell}{\store}, \ell_c}{t} \quad \ell_c\Flows \ell\\
  \config{\hole,\hole}{\hole} \quad\text{otherwise}
  \end{cases}
\end{align*}

We say that two stores are \emph{$\ell$-equivalent} ($\store_1 \lequiv{\ell} \store_2$)
iff $\erase{\ell}{\store_1} = \erase{\ell}{\store_2}$.
Similarly, we say that two instrumented configurations are \emph{$\ell$-equivalent} ($k_1 \lequiv{\ell} k_2$)
iff $\erase{\ell}{k_1} = \erase{\ell}{k_2}$

\begin{lemma}[Noninterference of instrumented semantics] \label{appendix:lio}
If $k_1 \lequiv{\ell} k_2$ and $k_1 \liobigeval k_1'$ and $k_2 \liobigeval k_2'$,
then $k_1' \lequiv{\ell} k_2'$
\end{lemma}
This lemma follows directly from Theorem 1 in~\cite{StefanMMR17},
if we implement \kwget, \kwset, and \kwdown as specified in \autoref{sec:formal}.

\custompar{Simulation.}
We now prove that instrumented semantics simulates original semantics for well-typed closed terms.
We first prove this for pure term (terms of types other than \T{TIO}),
and then for \T{TIO} terms.

\begin{lemma}[Pure Simulation] \label{appendix:pure-simulation}
If $\epsilon\vdash t :: T$ where $T = \{B \mid r\}$, $T = \funT{x}{T_1}{T_2}$, or $T = \fld{T}{l}$,
and $\config{\store}{t} \bigeval \config{\store'}{v}$,
then for any $\ell_c$:
\begin{enumerate} 
\item\label{goal:same-pure} $\store' = \store$
\item\label{goal:steps-pure} $\config{\store,\instr{\ell_c}}{t} \liobigeval \config{\store,\instr{\ell_c}}{v}$,
\item\label{goal:checks-pure} $\epsilon\vdash v :: T$.
\end{enumerate}
\end{lemma}

\begin{proof}
By induction on $\config{\store}{t} \bigeval \config{\store}{v}$.
The only applicable cases are \textsc{val}, \textsc{app}, \textsc{if-true}, \textsc{if-false}; 
the rest of the cases contradict the premise that $t$ is well-typed at a non-\T{TIO} type.

\item \textbf{Case \textsc{val}:}
$$
  \inference[\textsc{val}]
    {}
    {\config{\store}{v} \bigeval \config{\store}{v}}
$$
Trivial by rule \textsc{i-val}.

\item \textbf{Case \textsc{app}:}
$$
  \inference[\textsc{app}]
    {\config{\store}{t_1} \bigeval \config{\store}{\lambda x.~ t_1'} & \config{\store}{[x\,{\mapsto}\,t_2]t_1'} \bigeval \config{\store'}{v}}
    {\config{\store}{t_1 ~ t_2} \bigeval \config{\store'}{v}}
$$
By rule \textsc{t-app}, $\epsilon\vdash t_1:: T' \to T$;
hence we can apply the IH to the first premise to get:
$\config{\store,\instr{\ell_c}}{t_1} \liobigeval \config{\store,\instr{\ell_c}}{\lambda x.~ t_1'}$ \fact{a}.
By \textsc{t-app} and substitution lemma for refinement types from prior work (\eg~\cite{KnowlesF10}), 
$\epsilon\vdash [x\,{\mapsto}\,t_2]t_1' :: T$;
hence we can apply the IH also to the second premise to get:
$\store' = \store$ (hence \ref{goal:same-pure} holds), 
$\config{\store,\instr{\ell_c}}{[x\,{\mapsto}\,t_2]t_1'} \liobigeval \config{\store,\instr{\ell_c}}{v}$ \fact{b},
and $\epsilon\vdash v::T$ (hence \ref{goal:checks-pure} holds).
From \fact{a} and \fact{b} we conclude \ref{goal:steps-pure} by rule \textsc{i-app}.

\item \textbf{Case \textsc{if-true}:}
$$
  \inference[\textsc{if-true}]
    {\config{\store}{t} \bigeval \config{\store}{\vtrue} & \config{\store}{t_1}\bigeval\config{\store'}{v}}
    {\config{\store}{\kwif ~ t ~ \kwthen ~ t_1 ~ \kwelse ~ t_2} \bigeval \config{\store'}{v}}
$$
From the typing premise by rule \textsc{t-if}, 
$\epsilon \vdash t :: \{\T{Bool} \mid r\}$ and $\epsilon, [\nu\mapsto\true]r \vdash t_1::T$ \fact{a}.
We can apply the IH to the first premise to get:
$\config{\store,\instr{\ell_c}}{t} \liobigeval \config{\store,\instr{\ell_c}}{\vtrue}$ \fact{b}
and $\epsilon\vdash \vtrue ::\{\T{Bool} \mid r\}$.
By soundness of refinement types, the latter implies that $[\nu\mapsto\true]r \liff \true$;
hence we can remove the trivial path constraint from the environment in \fact{a} and get $\epsilon \vdash t_1::T$.
With that we can apply the IH to the second premise, obtaining:
$\store' = \store$ (hence \ref{goal:same-pure} holds), 
$\config{\store,\instr{\ell_c}}{t_1} \liobigeval \config{\store,\instr{\ell_c}}{v}$ \fact{c},
and $\epsilon\vdash v::T$ (hence \ref{goal:checks-pure} holds).
From \fact{b} and \fact{c} we conclude \ref{goal:steps-pure} by rule \textsc{i-if-true}.

\item \textbf{Case \textsc{if-false}:}
Symmetric to \textsc{if-true}.
\end{proof}

\begin{lemma}[Simulation] \label{appendix:simulation}
If $\epsilon\vdash t :: \tio{T}{\ell_i}{\ell_o}$ and $\config{\store}{t} \bigeval \config{\store'}{v}$,
then for any $\ell_c \Flows \sem{\ell_o}$, there exists a new current label $\ell_c'$ such that
\begin{enumerate} 
\item\label{goal:steps} $\config{\store,\instr{\ell_c}}{t} \liobigeval \config{\store',\instr{\ell'_c}}{v}$,
\item\label{goal:input} $\ell_c' \Flows \ell_c \Fjoin \sem{\ell_i}$,
\item\label{goal:checks} $\epsilon\vdash v :: \tio{T}{\ell_i}{\ell_o}$.
\end{enumerate}
\end{lemma}
\begin{proof}
By induction on the derivation of $\config{\store}{t} \bigeval \config{\store'}{v}$.

\item \textbf{Case \textsc{get}:}
$$
  \inference[\textsc{get}]
    {\config{\store}{t} \bigeval \config{\store}{f}  \quad\text{\fact{a}}}
    {\config{\store}{\kwget ~ t} \bigeval \config{\store}{\tioval{\store[f]}}}
$$
From the well-typing premise we know $\epsilon \vdash \kwget ~ t :: \tio{T}{\ell}{\top}$ and hence $\epsilon \vdash t :: \fld{T}{\ell}$ \fact{b} by rule \textsc{t-get}.
Let $\ell_c$ be arbitrary and the new current label be $\ell_c' = \ell \Fjoin\ell_c$.
From \fact{a} and \fact{b} by pure simulation (Lemma~\ref{appendix:pure-simulation}),
we know that $\config{\store,\instr{\ell_c}}{t} \bigeval \config{\store,\instr{\ell_c}}{f}$ \fact{c} and $\epsilon \vdash f :: \fld{T}{\ell}$,
and hence $\labelof(f) = \ell$ \fact{d} by \textsc{t-Field} and because field types are invariant in the label.
\begin{itemize}
\item From \fact{c} and \fact{d} by rule \textsc{i-get}: 
$\config{\store,\instr{\ell_c}}{\kwget ~ t} \liobigeval \config{\store,\instr{\ell\Fjoin \ell_c}}{\tioval{\store[f]}}$, hence (\ref{goal:steps}) holds.
\item $\ell \Fjoin \ell_c \Flows \ell \Fjoin \ell_c$, hence (\ref{goal:input}) holds.
\item By rule \textsc{t-TIO}: $\epsilon \vdash \tioval{\store[f]} :: \tio{T}{\bot}{\top} \Subt \tio{T}{\ell}{\top}$, hence (\ref{goal:checks}) holds.
\end{itemize}

\item \textbf{Case \textsc{set}:}
$$
  \inference[\textsc{set}]
    {\config{\store}{t_1} \bigeval \config{\store}{f} & \config{\store}{t_2} \bigeval \config{\store}{b}}
    {\config{\store}{\kwset ~ t_1 ~ t_2} \bigeval \config{\store[f := b]}{\tioval{\unit}}}  
$$
By well-typing premise and rule \textsc{t-set}, we have 
$\epsilon \vdash \kwset ~ t_1 ~ t_2 :: \tio{\T{()}}{\bot}{\ell}$ and hence $\epsilon \vdash t_1 :: \fld{T}{\ell}$ and $\epsilon \vdash t_2 :: T$.
Pick some $\ell_c \Flows \ell$.
By pure simulation for $t_1$ we get $\config{\store,\instr{\ell_c}}{t} \bigeval \config{\store,\instr{\ell_c}}{f}$ \fact{a}
and $\epsilon \vdash f :: \fld{T}{\ell}$ and hence $\labelof(f) = \ell$ \fact{b}.
Because $T$ must be a (refined) base type, 
by pure simulation for $t_2$ we we get $\config{\store,\instr{\ell_c}}{t} \bigeval \config{\store,\instr{\ell_c}}{b}$ \fact{c} and $\epsilon\vdash b :: T$.
Let the new current label be unchanged, \ie $\ell_c' = \ell_c$.
\begin{itemize}
\item From \fact{b} we know that $\ell_c \Flows \sem{\labelof(f)}$, hence by rule \textsc{i-set} from \fact{a} and \fact{c}, we have $\config{\store,\instr{\ell_c}}{\kwset ~ t_1 ~ t_2} \liobigeval \config{\store[f := b],\instr{\ell_c}}{\tioval{\unit}}$, hence (\ref{goal:steps}) holds.
\item $\ell_c \Flows \ell_c \Fjoin \bot$, hence (\ref{goal:input}) holds.
\item By rule \textsc{t-TIO}: $\epsilon \vdash \tioval{\unit} :: \tio{\T{()}}{\bot}{\top} \Subt \tio{\T{()}}{\bot}{\ell}$, hence (\ref{goal:checks}) holds.
\end{itemize}

\item \textbf{Case \textsc{ret}:}
$$
  \inference[\textsc{ret}]
    {}
    {\config{\store}{\kwreturn~t} \bigeval \config{\store}{\tioval{t}}}
$$
By the well-typing premise $\epsilon \vdash \kwreturn ~ t :: \tio{T}{\bot}{\top}$, and hence $\epsilon \vdash t :: T$  \fact{a} by rule \textsc{t-ret}.   
Pick any $\ell_c$, and let the new current label be unchanged, \ie $\ell_c' = \ell_c$.
\begin{itemize}
\item By rule \textsc{i-ret}: $\config{\store,\instr{\ell_c}}{\kwreturn~t} \liobigeval \config{\store,\instr{\ell_c}}{\tioval{t}}$, hence (\ref{goal:steps}) holds.
\item $\ell_c \Flows \ell_c \Fjoin \bot$, hence (\ref{goal:input}) holds.
\item By \fact{a} and rule \textsc{t-TIO}: $\epsilon \vdash \tioval{t} :: \tio{T}{\bot}{\top}$, hence (\ref{goal:checks}) holds.
\end{itemize}

\item \textbf{Case \textsc{bind}:}
$$
  \inference[\textsc{bind}]
    {\config{\store}{t_1} \bigeval \config{\store'}{\tioval{t_1'}} \quad\text{\fact{a}} &
      \config{\store'}{t_2 ~ t_1'} \bigeval \config{\store''}{v} \quad\text{\fact{b}}}
    {\config{\store}{\kwbind ~ t_1 ~ t_2} \bigeval \config{\store''}{v}}
$$
By the well-typing premise we have $\epsilon \vdash \kwbind ~ t_1 ~ t_2 :: \tio{T_2}{\ell_1 \Fjoin \ell_2}{\ell_1' \Fmeet \ell_2'}$,
and hence $\epsilon \vdash t_1 :: \tio{T_1}{\ell_1}{\ell_1'}$ \fact{c},
$\epsilon \vdash t_2 :: \funT{x}{T_1}{\tio{T_2}{\ell_2}{\ell_2'}}$ \fact{d},
and $\epsilon \vdash \ell_1 \Flows \ell_2'$ \fact{e}. 
Pick some $\ell_c \Flows \sem{\ell_1'} \Fmeet \sem{\ell_2'}$ \fact{f}.

We can apply the IH to \fact{a} and \fact{c} because $\ell_c \Flows \sem{\ell_1'}$.
Hence there exists $\ell_c'$ such that $\config{\store,\instr{\ell_c}}{t_1} \liobigeval \config{\store',\instr{\ell'_c}}{\tioval{t_1'}}$ \fact{g},
$\ell_c' \Flows \ell_c \Fjoin \sem{\ell_1}$ \fact{h},
and $\epsilon\vdash t_1' :: T_1$ \fact{i} (using rule \textsc{t-TIO}).

From \fact{d} and \fact{i} by rule \textsc{t-app}, we get 
$\epsilon \vdash t_2 ~ t_1' :: \tio{T_2}{\ell_2}{\ell_2'}$.
Now we would like to apply the IH to this and \fact{b} with the current label $\ell_c'$,
but we have to show that $\ell_c' \Flows \sem{\ell_2'}$.
To this end, we calculate: 
$$
\ell_c' \quad\ontop{\Flows}{\fact{h}}\quad
\ell_c \Fjoin \sem{\ell_1} \quad\ontop{\Flows}{\fact{e}}\quad
\ell_c \Fjoin \sem{\ell_2'} \quad\ontop{\Flows}{\fact{f}}\quad
((\sem{\ell_1'} \Fmeet \sem{\ell_2'}) \Fjoin \sem{\ell_2'}) \quad=\quad \sem{\ell_2'}
$$
Applying the IH, we get a new current label $\ell''_c$,
such that:
$\config{\store',\instr{\ell'_c}}{t_2 ~ t_1'} \liobigeval \config{\store'',\instr{\ell''_c}}{v}$ \fact{j},
$\ell''_c \Flows \ell'_c \Fjoin \sem{\ell_2}$ \fact{k},
$\epsilon\vdash v :: \tio{T_2}{\ell_2}{\ell_2'}$ \fact{l}.

Let the new current label be $\ell''_c$:
\begin{itemize}
\item By rule \textsc{i-bind} and \fact{g}, \fact{j} we get: 
$\config{\store,\instr{\ell_c}}{\kwbind ~ t_1 ~ t_2} \liobigeval \config{\store'',\instr{\ell''_c}}{v}$, hence (\ref{goal:steps}) holds.
\item To show (\ref{goal:input}) we calculate:
$$
\ell''_c \quad\ontop{\Flows}{\fact{k}}\quad
\ell'_c \Fjoin \sem{\ell_2} \quad\ontop{\Flows}{\fact{h}}\quad
(\ell_c \Fjoin \sem{\ell_1}) \Fjoin \sem{\ell_2} \quad=\quad 
\ell_c \Fjoin (\sem{\ell_1} \Fjoin \sem{\ell_2})
$$
\item (\ref{goal:checks}) follows immediately by \fact{l}.
\end{itemize}

\item \textbf{Case \textsc{down}:}
$$
  \inference[\textsc{down}]
    {\config{\store}{t_1} \bigeval \config{\store}{f} \quad\text{\fact{a}} & 
     \config{\store}{t_2} \bigeval \config{\store'}{\tioval{t}} \quad\text{\fact{b}} & 
     \config{\store'}{t} \bigeval \config{\store'}{b}  \quad\text{\fact{c}}}
    {\config{\store}{\kwdown~t_1~t_2} \bigeval \config{\store'}{\tioval{b}}}  
$$    
By well-typing premise, we have $\epsilon \vdash \kwdown ~ t_1 ~ t_2 :: \tio{\T{Bool}}{\ell}{\ell'}$
and hence $\epsilon \vdash t_1 :: \fld{\_}{\ell}$ \fact{d} and $\epsilon \vdash t_2 :: \tio{\{\T{Bool} \mid \nu \Implies r\}}{\ell \Fjoin r}{\ell'}$ \fact{e}.
Note that for this type to be well-formed, by rule \textsc{wf-TIO}
$r$ can mention neither $\nu$ nor $\upsilon$ (nor any other variable);
since it is a closed Boolean-sorted refinement term,
we know that either $r \liff \true$ or $r \liff \false$.
Now pick some $\ell_c \Flows \sem{\ell'}$.

From \fact{a} and \fact{d} by pure simulation we get 
$\config{\store,\instr{\ell_c}}{t_1} \liobigeval \config{\store,\instr{\ell_c}}{f}$\fact{f}
and $\epsilon \vdash f :: \fld{\_}{\ell}$, so $\labelof(f) = \ell$ \fact{g}.
We can apply the IH to \fact{b} and \fact{e} to obtain a label $\ell'_c$ such that:
$\config{\store,\instr{\ell_c}}{t_2} \liobigeval \config{\store',\instr{\ell'_c}}{\tioval{t}}$ \fact{h},
$\ell_c' \Flows \ell_c \Fjoin \sem{\ell} \Fjoin \sem{r}$ \fact{i},
and $\epsilon\vdash t :: \{\T{Bool} \mid \nu \Implies r\}$ (using \textsc{t-TIO}).
From this and \fact{c} by pure simulation we get
$\config{\store',\instr{\ell_c'}}{t} \liobigeval \config{\store',\instr{\ell_c'}}{b}$\fact{j}
and $\epsilon \vdash b :: \{\T{Bool} \mid \nu \Implies r\}$\fact{k}.

As the new current label we pick $\sem{\ell}\Fjoin \ell_c$.
\begin{itemize}
\item To prove that the downgrade steps to the same term under the instrumented semantics, we consider two cases.
First, if $\ell_c' \Flows \sem{\ell}\Fjoin \ell_c$, then from \fact{f}, \fact{g}, \fact{h}, and \fact{j} by rule \textsc{i-down-1}, we have 
$\config{\store,\instr{\ell_c}}{\kwdown~t_1~t_2} \liobigeval \config{\store',\instr{\sem{\ell}\Fjoin \ell_c}}{\tioval{b}}$, hence (\ref{goal:steps}) holds trivially in this case.

Otherwise, by rule \textsc{i-down-2}, we have
$\config{\store,\instr{\ell_c}}{\kwdown~t_1~t_2} \liobigeval \config{\store',\instr{\sem{\ell}\Fjoin \ell_c}}{\tioval{\vfalse}}$.
So it suffices to show that $b = \vfalse$ in this case, then (\ref{goal:steps}) will follow.
From \fact{k} and typing rules we know that $b$ is necessarily either \vtrue or \vfalse;
so let us assume $b = \vtrue$ and arrive at a contradiction.
If $b = \vtrue$, then by rule \textsc{t-True}, $\epsilon\vdash b :: \{\T{Bool}\mid \nu\}$;
then from \fact{k} and subtyping we get $\epsilon\vdash \{\T{Bool}\mid \nu\} \Subt \{\T{Bool} \mid \nu \Implies r\}$,
and hence $\epsilon \vDash \nu\Implies r$.
Recall, however, that $r$ is a closed term, so the only way this implication can be valid is if $r \liff \true$,
in which case $\sem{r} = \bot$ and so $\ell_c \Fjoin \sem{\ell} \Fjoin \sem{r} = \ell_c \Fjoin \sem{\ell}$.
Hence from \fact{i} we get $\ell_c' \Flows \ell_c \Fjoin \sem{\ell}$,
which directly contradicts our assumption.

\item $\ell\Fjoin \ell_c \Flows \ell \Fjoin \ell_c$, hence (\ref{goal:input}) holds.
\item From \fact{k} by rule \textsc{t-TIO} and subtyping: $\epsilon\vdash \tioval{b} :: \tio{\T{Bool}}{\bot}{\top}$, hence (\ref{goal:checks}) holds.
\end{itemize}

\item \textbf{Case \textsc{val}:}
$$
  \inference[\textsc{val}]
    {}
    {\config{\store}{v} \bigeval \config{\store}{v}}
$$
Trivial by keeping the current label unchanged.

\item \textbf{Case \textsc{app}:}
$$
  \inference[\textsc{app}]
    {\config{\store}{t_1} \bigeval \config{\store}{\lambda x.~ t_1'}   \quad\text{\fact{a}} & 
    \config{\store}{[x\,{\mapsto}\,t_2]t_1'} \bigeval \config{\store'}{v} \quad\text{\fact{b}}}
    {\config{\store}{t_1 ~ t_2} \bigeval \config{\store'}{v}}
$$
By the well-typing premise we get $\epsilon \vdash t_1 ~ t_2 :: \tio{T}{\ell_i}{\ell_o}$     
and hence $\epsilon \vdash t_1 :: \funT{x}{T_2}{\tio{T}{\ell_i}{\ell_o}}$ \fact{c} and $\epsilon \vdash t_2::T_2$ \fact{d}.
Let us pick some $\ell_c \Flows \sem{\ell_o}$.
From \fact{a} and \fact{c} by Lemma~\ref{appendix:pure-simulation},
we get $\config{\store,\instr{\ell_c}}{t_1} \liobigeval \config{\store,\instr{\ell_c}}{\lambda x.~ t_1'}$ \fact{e}
and $\epsilon\vdash \lambda x.~ t_1' :: \funT{x}{T_2}{\tio{T}{\ell_i}{\ell_o}}$ \fact{f}.

From \fact{f} and \fact{d}, by substitution lemma for refinement types,
we get $\epsilon \vdash [x\,{\mapsto}\,t_2]t_1' :: \tio{T}{\ell_i}{\ell_o}$.
We can apply the IH to this fact and \fact{b} with current label $\ell_c$
to obtain a new current label $\ell'_c$, which we adopt as the new label for the whole application.
IH gives us:
\begin{itemize}
\item $\config{\store, \instr{\ell_c}}{[x\,{\mapsto}\,t_2]t_1'} \liobigeval \config{\store',\instr{\ell'_c}}{v}$, 
  which together with \fact{e} by rule \textsc{i-app} implies (\ref{goal:steps});
\item $\ell_c' \Flows \ell_c \Fjoin \sem{\ell_i}$, which proves (\ref{goal:input});
\item $\epsilon\vdash v :: \tio{T}{\ell_i}{\ell_o}$, which proves (\ref{goal:checks}).
\end{itemize}
\end{proof}

\custompar{Noninterference.}
Putting is all together, we obtain the following noninterference theorem:

\begin{theorem}[Noninterference for \corelg] \label{appendix:noninterference}
Take two $\ell$-equivalent stores $\store_1 \lequiv{\ell} \store_2$.
Then for any term $t$ such that $\epsilon\vdash t :: \tio{T}{\ell}{\_}$,
$\config{\store_1}{t} \bigeval \config{\store_1'}{v_1}$,
and $\config{\store_2}{t} \bigeval \config{\store_2'}{v_2}$,
we have $v_1 = v_2$ and $\store_1' \lequiv{\ell} \store_2'$.
\end{theorem}
\begin{proof}
Since $\config{\store_1}{t} \bigeval \config{\store_1'}{v_1}$
and $\epsilon\vdash t :: \tio{T}{\ell}{\_}$,
then by Lemma~\ref{appendix:simulation} we have
$\config{\store_1,\instr{\bot}}{t} \liobigeval \config{\store_1',\instr{\ell_c^1}}{v_1}$
and $\ell_c^1 \Flows \ell$.
Similarly for the other store we get:
$\config{\store_2,\instr{\bot}}{t}  \liobigeval \config{\store_2',\instr{\ell_c^2}}{v_2}$
and $\ell_c^2 \Flows \ell$.
Because 
$\config{\store_1,\instr{\bot}}{t} \lequiv{\ell} \config{\store_2,\instr{\bot}}{t}$,
then by Lemma~\ref{appendix:lio} we get:
$\config{\store_1',\instr{\ell_c^1}}{v_1} \lequiv{\ell} \config{\store_2',\instr{\ell_c^2}}{v_2}$.
Now since $\ell_c^1 \Flows \ell$ and $\ell_c^2 \Flows \ell$, then by definition of configuration erasure
we get that $\store_1' \lequiv{\ell} \store_2'$ and $v_1 = v_2$.
\end{proof}

\section{Scalability}\label{appendix:scalability}

\autoref{eval:benchmarks:scalability-OneFunc} shows the dependency of verification and patch generation times
on the number of leaky input actions to patch.

\begin{figure}
\pgfplotsset{
    scale only axis
}
\begin{tikzpicture}[>=latex,scale=.8,mark options={thick}]
	\begin{axis}[
	    height=4cm, width=8.25cm,
        axis y line*=left,
        ymin=.0, ymax=25,
        xmin=.5, xmax=16.5,
		ylabel={Time ({\it s}\,)},
		scaled x ticks=false, 
		ymajorgrids=true,
		legend style={at={(0.02,0.98)},anchor=north west}]
	\addplot[color=olive!70!white,ultra thick,mark=x,smooth] table[x=N,y=Localize] 
	   {data/scalability-OneFunc.txt};
	\addlegendentry{Localize}
	\addplot[color=blue!50!white,ultra thick,mark=square*,smooth] table[x=N,y=Generate]
	   {data/scalability-OneFunc.txt};
	\addlegendentry{Generate}
	
	\addplot[color=red!70!white,ultra thick,mark=*,smooth] table[x=N,y=Total] 
	   {data/scalability-OneFunc.txt};
	\addlegendentry{Total}	
	
	\end{axis}
	
	\node[right] at (8.5,0) {\small N};
\end{tikzpicture}
\vspace{-.3cm}
\caption{\label{eval:benchmarks:scalability-OneFunc}
   Scalability---N leaky input actions in a single function.}
\end{figure}

\section{Microbenchmarks}\label{appendix:benchmarks}
Below you can find the code for all our microbenchmarks with \toolname-generated patches in gray.

\newcommand\benchmarktitle[1]{\vspace{3pt}\noindent \textbf{#1}~~}
\newcommand\benchmarkname[1]{({\tt #1})}

\newpage
\benchmarktitle{Benchmark 1 \benchmarkname{EDAS}}: Show data for paper \T{p} to \T{client}.

{\small
\begin{lifty}
-- | Conference phase (public)
predicate phase :: Store -> Phase
getPhase :: ds: Store -> TIO {Phase | _v == phase ds} <{True}> <{False}>
-- | Paper title (public)
getPaperTitle :: ds: Store -> p: PaperId -> TIO String <{True}> <{False}>
-- | Paper status (only visible when phase is done)
getPaperDecision :: ds: Store -> p: PaperId -> TIO Decision <{phase ds == Done}> <{False}>
-- | Paper session (public)
getPaperSession :: ds: Store -> p: PaperId -> TIO String <{True}> <{False}>
redact {NoDecision}

showSession :: Store -> User -> PaperId -> TIO Unit <False> <True>
showSession = \ds . \client . \p . 
  do
     t <- getPaperTitle ds p
     dec <- do
              @g0 <- downgrade (do
                                 x5 <- getPhase ds
                                 return (eq Done x5))              
              if g0
                then @getPaperDecision ds p@
                else return NoDecision@           
     if dec == Accepted
       then do
              ses <- getPaperSession ds p
              print client (unwords [t, ses])
       else print client (unwords [t, emptyString])
\end{lifty}
}

\newpage
\benchmarktitle{Benchmark 2 \benchmarkname{EDAS-Multiple}}: Same as \benchmarkname{EDAS}, but multiple terms need to be patched.

{\small
\begin{lifty}
... -- as in EDAS
-- | Paper authors (only visible when phase is done)
getPaperAuthors :: ds: Store -> p: PaperId -> TIO [User] <{phase ds == Done}> <{False}>

showSession :: Store -> User -> PaperId -> TIO Unit <False> <True>
showSession = \ds . \client . \p . 
  do
     t <- getPaperTitle ds p
     auts <- do
               @g0 <- downgrade (do
                                  x5 <- getPhase ds
                                  return (eq Done x5))               
               if g0
                 then @getPaperAuthors ds p@
                 else return Nil@
     dec <- do
              @g1 <- downgrade (do
                                 x11 <- getPhase ds
                                 return (eq Done x11))                    
              if g1
                then @getPaperDecision ds p@
                else return NoDecision@           
     if dec == Accepted
       then do
              ses <- getPaperSession ds p
              print client (unwords [t, show auts, ses])
       else print client (unwords [t, show auts, emptyString])
\end{lifty}


\newpage
\benchmarktitle{Benchmark 3 \benchmarkname{EDAS-Self-Ref}}: Same as \benchmarkname{EDAS-Multiple}, 
but with a self-referential policy on \T{authors}.

{\small
\begin{lifty}
... -- as in EDAS
-- | Paper authors (only visible to themselves or when phase is done)
predicate paperAuthors :: Store -> Map PaperId (Set User)
getPaperAuthors :: ds: Store -> p: PaperId -> 
  TIO {List {User | _v in (paperAuthors ds)[[p]]} | elems _v == (paperAuthors ds)[[p]]} 
    <{_0 in (paperAuthors ds)[[p]] || phase ds == Done}> <{False}>
    
showSession :: Store -> User -> PaperId -> TIO Unit <False> <True>
showSession = \ds . \client . \p . 
  do
     t <- getPaperTitle ds p
     auts <- do
               @g0 <- downgrade (do
                                  x5 <- getPhase ds
                                  return (eq Done x5))
               g1 <- downgrade (do
                                  x10 <- getPaperAuthors ds p
                                  return (elem client x10))                     
               if g0 || g1
                 then @getPaperAuthors ds p@
                 else return Nil@
     dec <- do
              @g2 <- downgrade (do
                                 x16 <- getPhase ds
                                 return (eq Done x16))              
              if g2
                then @getPaperDecision ds p@
                else return NoDecision@           
     if dec == Accepted
       then do
              ses <- getPaperSession ds p
              print client (unwords [t, show auts, ses])
       else print client (unwords [t, show auts, emptyString])
\end{lifty}


\newpage
\benchmarktitle{Benchmark 4 \benchmarkname{Search}}: Show client all their accepted papers. Repairs a leak through a filter.
          
{\small
\begin{lifty}
... -- as in EDAS-Self-Ref

showMyAcceptedPapers :: Store -> User -> TIO Unit <False> <True>
showMyAcceptedPapers = \ds .\client . 
    let isMyAccepted = \p .
        downgrade (do
                     auts <- getPaperAuthors ds p
                     dec <- do
                              @g0 <- downgrade (do
                                                 x5 <- getPhase ds
                                                 return (eq Done x5))                              
                              if g0
                                then @getPaperDecision ds p@
                                else return NoDecision@
                     return ((elem client auts) &&
                               (dec == Accepted))) in
    do
      allPaperIDs <- getAllPaperIds ds
      paperIDs <- filterM isMyAccepted
                    allPaperIDs
      titles <- mapM (\p . getPaperTitle ds p) paperIDs
      print client (unlines titles)
\end{lifty}
}

\newpage
\benchmarktitle{Benchmark 5 \benchmarkname{Sort}} Sort papers by their score, which is hidden from conflicted reviewers. 
Repairs a leak through the order of sorted submission. Contains a negative self-referential policy. 
          
{\small
\begin{lifty}
-- | Paper title (public)
getPaperTitle :: ds: Store -> p: PaperId -> TIO String <{True}> <{False}>
-- | Paper conflicts (public)
predicate paperConflicts :: Store -> Map PaperId (Set User)
getPaperConflicts :: ds: Store -> pid: PaperId 
  -> TIO {List User | elems _v == (paperConflicts ds)[[pid]]} 
          <{!(_0 in (paperConflicts ds)[[pid]])}> <{False}>
-- | Paper score (only visible if not in conflict)
getPaperScore :: ds: Store -> pid: PaperId -> TIO Int 
          <{!(_0 in (paperConflicts ds)[[pid]])}> <{False}>
-- | All papers in the conference
getAllPaperIds :: ds: Store -> TIO [PaperId] <{True}> <{False}>

sortPapersByScore :: Store -> User -> TIO Unit <False> <True>
sortPapersByScore = \ds . \client . 
  let cmpScore = \pid1 . \pid2 .
        do
          x1 <- do
                  @g0 <- downgrade (do
                                     x7 <- getPaperConflicts ds pid1
                                     return (not (elem client x7)))                  
                  if g0
                    then @getPaperScore ds pid1@
                    else return zero@
          x2 <- do
                  @g1 <- downgrade (do
                                     x15 <- getPaperConflicts ds pid2
                                     return (not (elem client x15)))                  
                  if g1
                    then @getPaperScore ds pid2@
                    else return zero@
          return (x1 <= x2) in
  do
    pids <- getAllPaperIds ds
    sortedPids <- sortByM cmpScore
                    pids
    titles <- mapM (getPaperTitle ds) sortedPids
    print client (unlines titles)
\end{lifty}
}

\newpage
\benchmarktitle{Benchmark 6 \benchmarkname{Broadcast}}: Send status notification to authors.
Sending message to multiple viewers, viewers are sensitive.

{\small
\begin{lifty}
-- | Conference phase (public)
predicate phase :: Store -> Phase
getPhase :: ds: Store -> TIO {Phase | _v == phase ds} <{True}> <{False}>
-- | Paper title (public)
getPaperTitle :: ds: Store -> p: PaperId -> TIO String <{True}> <{False}>
-- | Paper authors (only visible to themselves or when phase is done)
predicate paperAuthors :: Store -> Map PaperId (Set User)
getPaperAuthors :: ds: Store -> p: PaperId -> 
  TIO {List {User | _v in (paperAuthors ds)[[p]]} | elems _v == (paperAuthors ds)[[p]]} 
    <{_0 in (paperAuthors ds)[[p]] || phase ds == Done}> <{False}>    
-- | Paper status (only visible when phase is done)
getPaperDecision :: ds: Store -> p: PaperId -> TIO Decision <{phase ds == Done}> <{False}>
-- | Paper session (public)
getPaperSession :: ds: Store -> p: PaperId -> TIO String <{True}> <{False}>
redact {NoDecision}

notifyAuthors :: ds:Store -> p:PaperId -> TIO Unit <False> <True>
notifyAuthors = \ds . \p . 
  do
     status <- do
                 @g0 <- downgrade (do
                                    x5 <- getPhase ds
                                    return (eq Done x5))
                 
                 if g0
                   then @getPaperDecision ds p@
                   else return NoDecision@
     authors <- getPaperAuthors ds p
     printAll authors (show status)
\end{lifty}
}

\newpage
\benchmarktitle{Benchmark 7 \benchmarkname{HotCRP}}: HotCRP password leak: chair could see other people's passwords in message preview.

{\small
\begin{lifty}
-- | Mask a password
mask :: TIO Password <{False}> <{False}> -> TIO Password <{True}> <{False}>
-- | User name (public)
getUserName :: ds: Store -> u: User -> TIO String <{True}> <{False}>
-- | User password (only visible to the user)   
getUserPassword :: ds: Store -> u: User -> TIO Password <{_0 == u}> <{False}>
-- | PC chair (public)
predicate chair :: Store -> User
getChair :: ds: Store -> TIO {User | _v == chair ds} <{True}> <{False}>
redact {mask}

sendPasswordReminder :: Store -> User -> TIO Unit <False> <True>
sendPasswordReminder = \ds . \u . 
  do
     ch <- getChair ds
     preview <- liftM2 strcat
                  (getUserName ds u) (liftM show
                                        (do
                                           @g0 <- downgrade (return (eq ch
                                                                      u))
                                           
                                           if g0
                                             then @getUserPassword ds u@
                                             else mask (getUserPassword ds u)))@
     print ch preview
     message <- liftM2 strcat
                  (getUserName ds u) (liftM show (getUserPassword ds u))
     print u message
\end{lifty}
}

\newpage
\benchmarktitle{Benchmark 8 \benchmarkname{AirBnB}}
AirBnB bug: they scrape phone numbers from user messages, but forgot to do so in the preview.
  This example features a custom redaction function that makes a message text visible to the recipient,
  but not completely public.
  It also showcases expressive functional reasoning with higher-order functions,
  since correctness depends on the argument of \T{filterM}.

{\small
\begin{lifty}
getAllMessageIDs :: ds: Store -> [MessageId]
-- | AirBnB admin
predicate admin :: Store -> User
getAdmin :: ds: Store -> TIO {User | _v == admin ds} <{True}> <{False}>
-- | Message sender
predicate sender :: Store -> Map MessageId User
getSender :: ds: Store -> m: MessageId 
  -> TIO {User | _v == (sender ds)[[m]] && _v != (recipient ds)[[m]]} <{True}> <{False}>
-- | Message recipient
predicate recipient :: Store -> Map MessageId User
getRecipient :: ds: Store -> m: MessageId 
  -> TIO {User | _v == (recipient ds)[[m]] && _v != (sender ds)[[m]]} <{True}> <{False}>
-- | Message text (only visible to the admin and the sender)
getText :: ds: Store -> m: MessageId -> TIO String <{_0 == admin ds || _0 == (sender ds)[[m]]}> <{False}>
-- | Scrape phone numbers from the message, making it visible to the recipient
scrapePhoneNumbers :: ds: Store -> m: MessageId 
  -> TIO String <{_0 == admin ds || _0 == (sender ds)[[m]]}> <{False}>
  -> TIO String <{_0 == admin ds || _0 == (sender ds)[[m]] || _0 == (recipient ds)[[m]]}> <{False}>
redact {scrapePhoneNumbers}

viewInbox :: Store -> User -> TIO Unit <False> <True>
viewInbox = \ds . \client . 
  let isMyMessage = \m . do
                           to <- getRecipient ds m
                           return (to == client) in
  do
    myMIDs <- filterM isMyMessage
                (getAllMessageIDs ds)
    messages <- mapM (\m . do
                             @g0 <- downgrade (do
                                                x37 <- getAdmin ds
                                                return (eq client x37))
                             
                             if g0
                               then @getText ds m@
                               else scrapePhoneNumbers ds m (getText ds m)) myMIDs@
    print client (unlines messages)
\end{lifty}
}

\newpage
\benchmarktitle{Benchmark 9 \benchmarkname{Instagram}}: The James Comey Instagram leak: the follow-relationships of private accounts leak through recommendation algorithms.

{\small
\begin{lifty}
getAllUsers :: ds: Store -> [User]
-- | User name (visible to all)
predicate name :: Store -> Map User String
getName :: ds: Store -> u: User -> TIO {String | _v == (name ds)[[u]]} <{True}> <{False}>
-- | Is user's account private? (visible to all)
predicate isPrivate :: Store -> Map User Bool
getIsPrivate :: ds: Store -> u: User -> TIO {Bool | _v == (isPrivate ds)[[u]]} <{True}> <{False}>
-- | Who this user follows (for private accounts: only visible to followers)
predicate following :: Store -> Map User (Set User)
getIsFollowing :: ds: Store -> who: User -> whom: User ->
  TIO {Bool | _v == (whom in (following ds)[[who]])} 
    <{canSee ds _0 who && canSee ds _0 whom}> <{False}>
setIsFollowing :: ds: Store -> who: User -> whom: User -> val: Bool ->
  TIO {Store | (whom in (following _v)[[who]]) == val && name _v == name ds && isPrivate _v == isPrivate ds} 
    <{True}> <{canSee ds _0 who && canSee ds _0 whom}>
-- | Is account u public? Yes if it's not private    
inline isPublic ds u = !(isPrivate ds)[[u]]    
-- | Can x see y? Yes if they are the same, y is public, or x is following y 
inline canSee ds x y  =  x == y || isPublic ds y || y in (following ds)[[x]]

showRecommendations :: ds:Store -> client:User 
  -> newFriend:{User|isPublic ds _v && _v in (following ds)[[client]]} -> TIO Unit <False> <True>
showRecommendations = \ds .\client . \newFriend . 
  do
     uids <- 
       let shouldRecommend = \u1 . 
             do
               alreadyFollowing <- downgrade (getIsFollowing ds client u1)               
               if (u1 == client) || alreadyFollowing
                 then return false
                 else do
                        @g0 <- downgrade
                            (do
                               x5 <- getIsPrivate ds u1
                               return (not x5))
                        g1 <- downgrade (getIsFollowing ds client u1)                        
                        if g0 || g1
                          then @getIsFollowing ds newFriend u1@
                          else return false@
         in filterM (\u . shouldRecommend u) (getAllUsers ds)
     names <- mapM (\u . getName ds u) uids
     print client (unlines names)

follow :: ds:Store -> client:User -> newFriend:{User|isPublic ds _v} -> TIO Unit <False> <True>
follow = \ds . \client . \newFriend . 
  do
     ds' <- setIsFollowing ds client newFriend true
     showRecommendations ds' client newFriend
\end{lifty}
}

\end{document}